\documentclass[twocolumn]{svjour3}         

\smartqed

\usepackage{amsmath}
\usepackage{graphicx}
\usepackage{fixltx2e}
\usepackage[mathscr]{eucal}

\begin{document}

\title{The method of unitary clothing transformations in the theory of nucleon--nucleon scattering}

\author{I. Dubovyk         \and
        O. Shebeko }

\institute{I. Dubovyk \at
              Institute of Electrophysics \& Radiation Technologies, \\
              NAS of Ukraine, Kharkov, Ukraine \\
              \email{e.a.dubovik@mail.ru}
           \and
           O. Shebeko \at
              NSC Kharkov Institute of Physics \& Technology \\
              NAS of Ukraine, Kharkov, Ukraine \\
              \email{shebeko@kipt.kharkov.ua}
}

\date{Received: date / Accepted: date}

\maketitle

\begin{abstract}
The clothing procedure, put forward in quantum field
theory (QFT) by Greenberg and Schweber, is applied for the
description of nucleon--nucleon ($N-N$) scattering. We consider
pseudoscalar ($\pi$ and $\eta$), vector ($\rho$ and $\omega$) and
scalar ($\delta$ and $\sigma$) meson fields interacting with
$1/2$ spin ($N$ and $\bar N$) fermion ones via the
Yukawa--type couplings to introduce trial interactions between
"bare" particles. The subsequent unitary clothing transformations
(UCTs) are found to express the total Hamiltonian through new
interaction operators that refer to particles with physical
(observable) properties, the so--called clothed particles. In this
work, we are focused upon the Hermitian and energy--independent
operators for the clothed nucleons, being built up in the second
order in the coupling constants. The corresponding analytic
expressions in momentum space are compared with the separate meson
contributions to the one--boson--exchange potentials in the meson
theory of nuclear forces. In order to evaluate the $T$ matrix of
the $N-N$ scattering we have used an equivalence theorem that
enables us to operate in the clothed particle representation (CPR)
instead of the bare particle representation (BPR) with its large
amount of virtual processes. We have derived the
Lippmann--Schwin\-ger type equation for the CPR elements of the $T-$ matrix
for a given collision energy in the two--nucleon sector of the
Hilbert space $\mathcal{H}$ of hadronic states.
\end{abstract}

\section{Introductory remarks and some recollections}

We know that there are a number of high precision, boson-exchange
models of the two nucleon force $V_{NN}$, such as Paris
\cite{Lacom80}, Bonn \cite{MachHolElst87}, Nijmegen
\cite{Stocks94}, Argonne \cite{WirStocksSchia95}, CD Bonn
\cite{Mach01} potentials and a modern family of covariant
one-boson-exchange (OBE) ones \cite{GrossStad08}.Note also
successful treatments based on chiral effective field theory
\cite{OrdoRayKolck94},\cite{EpelGloeMeiss00}, for a review see
\cite{Epel05}.

In this paper, we would like to draw attention to the first
application of unitary clothing transformations (UCTs)
\cite{SheShi01,KorCanShe07} in describing the nucleon--nucleon
scattering. Recall that such transformations $W$, being
aimed at the inclusion of the so--called cloud or persistent
effects, make it possible the transition from the bare--particle
representation (BPR) to the clothed--particle representation (CPR)
in the Hilbert space  $\mathcal{H}$ of meson--nucleon states. In
this way, a large amount of virtual processes induced with the
meson absorption/emission, the $N\overline{N}-$pair annihilation/production
and other cloud effects can be accumulated in the creation (destruction) operators $\alpha_{c}$ for the "clothed" (physical) mesons and nucleons. Such a bootstrap reflects the most significant distinction between the concepts of
clothed and bare particles.

In the course of the clothing procedure all the generators of the
Poincar\'{e} group get one and the same sparse structure on
$\mathcal{H}$ \cite{SheShi01}. Here we will focus upon one of
them, viz., the total Hamiltonian
\begin{equation}
H = H_{F}(\alpha) + H_{I}(\alpha) \equiv H(\alpha) \label{Eq.1}
\end{equation}
with
\begin{equation}
H_{I}(\alpha) = V(\alpha) + \mbox{mass and vertex counterterms},
\end{equation}
where free part $H_{F}(\alpha) \sim \alpha^{\dag} \alpha$ belongs
to the class [1.1], if one uses the terminology adopted in
\cite{SheShi01}, and interaction $V(\alpha)$ is a function of
creation (destruction) operators $\alpha^{\dag} (\alpha)$ in the
BPR, i.e., referred to bare particles with physical masses \cite{KorCanShe07}, where they have been introduced via the mass--changing Bogoliu\-bov--type UTs. To be more definite, let us consider fermions (nucleons and
antinucleons) and bosons ($\pi-$, $\eta-$, $\rho-$,
$\omega$-mesons, etc.) interacting via the Yukawa-type couplings
for scalar (s), pseudoscalar (ps) and vector (v) mesons. Then, as
seen from Appendix A, $V(\alpha)=V_s + V_{ps} +V_{\rm{v}}$ with
\begin{equation}
V_s  = g_s \int d \vec x \, \bar \psi (\vec x)\psi (\vec x) \varphi_s (\vec x) \label{Eq.3}
\end{equation}
\begin{equation}
V_{ps}  = ig_{ps} \int d \vec x \, \bar \psi (\vec x) \gamma _5 \psi (\vec x) \varphi_{ps} (\vec x)
\end{equation}
\begin{multline}
V_{\rm{v}}  =
\int d \vec x \, \left\{ g_{\rm{v}} \bar \psi (\vec x) \gamma_\mu  \psi (\vec x) \varphi_{\rm{v}}^\mu  (\vec x) \phantom{\frac{f_{\rm{v}}}{4m}} \right. \\
\shoveright{ \left. + \frac{f_{\rm{v}}}{4m} \bar \psi (\vec x) \sigma_{\mu \nu} \psi (\vec x) \varphi_{\rm{v}}^{\mu \nu} (\vec x) \right\} } \\
\shoveleft{ + \int d \vec x \, \left\{ \frac{g_{\rm{v}}^{2}}{2m_{\rm{v}}^2} \bar \psi (\vec x) \gamma_0 \psi (\vec x) \bar \psi (\vec x) \gamma_0 \psi (\vec x) \right. } \\
\left. + \frac{f_{\rm{v}}^{2}}{4m^2} \bar \psi (\vec x) \sigma_{0i} \psi (\vec x) \bar \psi (\vec x) \sigma_{0i} \psi (\vec x) \right\},
\label{Eq.5}
\end{multline}
where $ \varphi_{\rm{v}}^{\mu \nu } (\vec x) = \partial^\mu  \varphi_{\rm{v}}^\nu (\vec x) - \partial^\nu \varphi_{\rm{v}}^\mu (\vec x) $ the tensor of the vector field
included. The mass (vertex) counterterms are given by Eqs. (32)--(33) of Ref. \cite{KorCanShe07} (the difference
$V_{0}(\alpha)$ - $V(\alpha)$ where a primary interaction $V_{0}(\alpha)$ is derived from $V(\alpha)$ replacing the "physical" coupling constants by "bare" ones).

The corresponding set $\alpha$ involves operators $ a^{\dag}(a) $
for the bosons, $ b^{\dag}(b) $ for the nucleons and $ d^{\dag}(d)
$ for the antinucleons. Following a common practice, they appear
in the standard Fourier expansions of the boson fields
$\varphi_{b}$ and the fermion field $\psi$, though for our
purposes the use of such a representation is not compulsory (see,
e.g., Chapter 3 of the monograph \cite{WeinbergBook1995}). In any
case we have the free pion and fermion parts
\begin{multline}
H_F (\alpha) = \int d \vec k \, \omega_{\vec k} a^\dag  (\vec k) a (\vec k) \\
+ \int d \vec p \, E_{\vec p} \sum \limits_{\mu}
\left[ b^\dag (\vec p, \mu) b (\vec p, \mu) + d^\dag (\vec p, \mu) d (\vec p, \mu) \right]
\end{multline}
and the primary trilinear interaction
\begin{equation}
V(\alpha) \sim a b^{\dag} b + a b^{\dag} d^{\dag} + a d b + a d
d^{\dag} + H.c. \label{Voperatorstructure}
\end{equation}
with the three-legs vertices. Here $\omega_{\vec k}  = \sqrt
{m^2_{b} + \vec k^2 } \,\,$ ($E_{\vec p}  = \sqrt {m^2  + \vec
p^2}$) represents the pion (nucleon) energy with physical mass
$m_{b} (m)$ while $\mu$ denotes the fermion polarization index. It
may be the particle spin projection onto the quantization axis (
the particle momentum ) for the so--called canonical (
helicity--state ) basis ( see, e.g., Chapter 4 of Ref.
\cite{Gas66} ).

In the context, we have tried to draw parallels with that
field--theoretic background which has been employed in
boson--exchange models. First of all, we imply the approach by the
Bonn group \cite{MachHolElst87,Mach01}, where, following the idea
by Sch\"{u}tte \cite{Schutte}, the authors started from the total
Hamiltonian (in our notations),
\begin{equation}
H = H_{F}(\alpha) + V(\alpha)
\end{equation}
with the boson-nucleon interaction
\begin{equation}
V(\alpha) \sim a b^{\dag} b + H.c.
\end{equation}
For brevity, the contributions from other mesons are omitted.{\sloppy

}
Unlike Eq.(\ref{Voperatorstructure}) the antinucleonic degrees of freedom were
disregarded in Ref. \cite{MachHolElst87}. There (see also \cite{Gas66}) the transition matrix
$T(z)$ (in the two--nucleon space) was considered  in the framework of the
three-dimensional perturbation theory when handling the integral equation,
\begin{equation}
T_{NN}(z)=V_{NN}(z)+V_{NN}(z)(z-H_{N})^{-1}T_{NN}(z),
\end{equation}
where the energy--dependent "quasipotential" $V_{NN}(z)$
approximates a sum of all relevant non-iterative diagrams, $H_{N}$
the nucleon contribution to $H_{F}$. Such a potential has "the
unpleasant feature of being energy--dependent. This complicates
applications to nuclear structure physics considerably" (quoted from p.40 in \cite{MachHolElst87}). Therefore, further
simplifications are welcome (see, e.g., Refs.
\cite{MachHolElst87}, \cite{Mach01} ).{\sloppy

}
Along with our derivation of a Lippmann--Schwinger (LS) equation
for the T matrix of the $N-N$ scattering, we will demonstrate its
solutions to be compared with those by the Bonn group.

\section{Analytic expressions for the quasipotentials in momentum space}

As shown in \cite{SheShi01}, after eliminating the so-called bad
terms\footnote{By definition, they prevent the bare vacuum
$\Omega_{0}$ ($a | \Omega_0 \rangle = b | \Omega_0 \rangle =\ldots= 0$) and the bare one--particle states
$|1bare\rangle \equiv a^{\dag} | \Omega_0 \rangle$ ($b^{\dag} | \Omega_0 \rangle,\ldots$) to be $H$ eigenstates.}
from $V(\alpha)$ the primary Hamiltonian $H(\alpha)$ can be represented in the form,
\begin{equation}
H(\alpha ) = K_F (\alpha _c ) + K_I (\alpha _c ) \equiv K(\alpha_c )
\end{equation}
The free part of the new decomposition is determined by
\begin{multline}
K_F(\alpha_c ) = \int d \vec k \, \omega_{\vec k} a_c^\dag (\vec k) a_c (\vec k) \\
+ \int d \vec p \, E_{\vec p} \sum\limits_{\mu} \left[ b_c^\dag (\vec p, \mu) b_c (\vec p, \mu) +
d_c^\dag (\vec p, \mu) d_c (\vec p, \mu) \right]
\end{multline}
while $K_{I}$ contains only interactions responsible for physical
processes, these quasipotentials between the clothed particles,
e.g.,
\begin{multline}
K_I^{(2)} (\alpha _c ) = K(NN \to NN) + K(\bar N\bar N \to \bar N\bar N) \\
+ K(N\bar N \to N\bar N) + K(b N \to b N) + K(b \bar N \to b \bar N) \\
+ K(b b^{\prime}  \to N \bar N) + K(N \bar N \to b b^{\prime} )
\label{K(2)general_structure}
\end{multline}
%

A key point of the clothing procedure developed in \cite{SheShi01}
is to fulfill the following requirements:

i) The physical vacuum (the $H$ lowest eigenstate) must coincide with a new no--particle state
$\Omega $, i.e., the state that obeys the equations
\begin{equation}
a_c (\vec k) \left\vert \Omega \right\rangle = b_c (\vec p, \mu) \left\vert \Omega \right\rangle = d_c (\vec p, \mu) \left\vert \Omega \right\rangle =0,
\,\,\, \forall \,\, {\vec k,\,\vec p,\,\mu} \label{ClothedOperatorDefinition}
\end{equation}
$$
\left\langle \Omega  | \Omega \right\rangle =1.
$$
ii) New one-clothed-particle states $| \vec k \rangle_c \equiv
a_c^{\dag} (\vec k) \Omega $ etc. are the
eigenvectors both of $K_F$ and $K$,
\begin{equation}
K(\alpha_c)|\vec k \rangle_c=K_F(\alpha_c)|\vec k \rangle_c = \omega_k|\vec k \rangle_c
\end{equation}
\begin{equation}
K_I(\alpha_c)|\vec k \rangle_c=0
\end{equation}

iii) The spectrum of indices that enumerate the new operators must
be the same as that for the bare ones .

iv) The new operators $\alpha_{c}$ satisfy the same commutation
rules as do their bare counterparts $\alpha$, since the both sets
are connected to each other via the similarity transformation
\begin{equation}
\alpha_{c}=W^{\dag} \alpha W ,
\end{equation}
with a unitary operator $W$ to be obtained as in \cite{SheShi01}.

It is important to realize that operator $K(\alpha_{c})$ is the
same Hamiltonian $H(\alpha)$. Accordingly [10,11] the $N$--$N$
interaction operator in the CPR has the following structure:
\begin{equation*}
K(NN \rightarrow NN) = \sum\limits_b K_{b}(NN \rightarrow NN),
\end{equation*}
\begin{multline}
K_{b}(NN \rightarrow NN) = \int \sum \limits_\mu d \vec p'_1 \, d \vec p'_2 \, d \vec p_1 \, d \vec p_2 \\
\times V_b (1', 2' ;1, 2) b_c^\dag (1') b_c^\dag (2') b_c (1) b_c (2),
\label{Eq.18}
\end{multline}
where the symbol $\sum\limits_{\mu}$ denotes the summation over
the nucleon spin projections, $1=\{ \vec p_1, \mu_1 \}$, etc.

For our evaluations of the c--number matrices $V_{b}$ we have employed some
experience from Refs. \cite{SheShi01,KorCanShe07} to get in the
second order in the coupling constants (see Appendix A)
\begin{multline}
V_b (1',2';1,2) = \frac{1}{(2\pi )^3} \frac{m^2}{\sqrt{E_{\vec p'_1} E_{\vec p'_2} E_{\vec p_1} E_{\vec p_2}}} \\
\times \delta \left( \vec p'_1 + \vec p'_2  - \vec p_1  - \vec p_2 \right) v_b (1',2' ;1,2),
\label{Eq.19}
\end{multline}
\begin{multline}
v_s (1',2' ;1,2) \\
=  - \frac{g_s^2}{2} \bar u(\vec p'_1) u(\vec p_1 ) \frac{1}{(p_1  - p'_1 )^2 - m_s^2} \bar u(\vec p'_2) u(\vec p_2), \label{Eq.20}
\end{multline}
\begin{multline}
v_{ps} (1',2' ;1,2) \\
= \frac{g_{ps}^2}{2}\bar u(\vec p'_1) \gamma_5 u(\vec p_1) \frac{1}{(p_1 - p'_1 )^2 - m_{ps}^2} \bar u(\vec p'_2) \gamma_5 u(\vec p_2 ),
\label{Eq.21}
\end{multline}
\begin{multline}
v_{\rm{v}} (1',2' ;1,2) =  \frac12 \frac{1}{(p'_1 - p_1 )^2  - m_{\rm{v}}^2} \\
\times \left[ \bar u (\vec p'_1) \left\{ ( g_{\rm{v}} + f_{\rm{v}} ) \gamma_\nu - \frac{f_{\rm{v}}}{2m}(p'_1 + p_1)_\nu \right\} u(\vec p_1) \right.  \\
\shoveright{ \times \bar u (\vec p'_2) \left\{ ( g_{\rm{v}} + f_{\rm{v}} ) \gamma^\nu - \frac{f_{\rm{v}}}{2m}(p'_2 + p_2)^\nu \right\} u(\vec p_2) } \\
\shoveright{ - \bar u (\vec p'_1) \left\{ (g_{\rm{v}} + f_{\rm{v}}) \gamma_\nu - \frac{f_{\rm{v}}}{2m}(p'_1  + p_1)_\nu \right\} u(\vec p_1) } \\
\times \bar u (\vec p'_2) \frac{f_{\rm{v}}}{2m} \left\{ (\hat p_1^{\prime}+\hat p_2^{\prime}-\hat p_1-\hat p_2)\gamma ^\nu  \right. \\
\left. \phantom{\frac{f_{\rm{v}}}{2m}} \left.  -(p'_1+p'_2-p_1-p_2)^{\nu} \right\} u( \vec p_2) \right], \label{Eq.22}
\end{multline}
where $m_{b}$ the mass of the clothed boson (its physical value) and $\hat q = q_{\mu}\gamma^{\mu}$. In the framework of the isospin formalism one needs to add the factor $\vec \tau (1) \vec \tau (2)$ in the corresponding expressions.

One should stress that in the course of our derivations the Feynman propagator
\begin{equation*}
\left[(p_{1}-p_{1}')^{2}-m^{2}_{b}\right]^{-1}
\end{equation*}
arises from adding the noncovariant propagators
\begin{equation*}
\left[2\omega_{\vec k} \left( E_{\vec p_1} - E_{\vec p'_1} - \omega_{\vec k} \right) \right]^{-1}
\end{equation*}
and
\begin{equation*}
\left[2\omega_{\vec k} \left( E_{\vec p'_1} - E_{\vec p_1} - \omega_{\vec k} \right) \right]^{-1}.
\end{equation*}
Such a feature of the UCTs method allows us to use the graphic language of the
old-fashioned perturbation theory (OFPT) (see, e.g., Chapter 13 in
Schweber's book \cite{SchweberBook}) when addressing the graphs in Fig.1.
\begin{figure}[h]
  \includegraphics[scale=0.55]{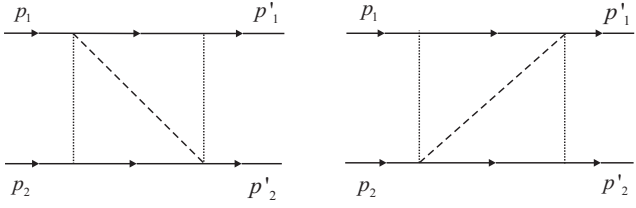}

\caption{The typical OFPT diagrams with the intermediate boson (dashed lines) on its mass shell.}
\label{fig:1}
\end{figure}
As noted in \cite{KorCanShe07} the graphs within our approach
should not be interpreted as the two time-ordered Feynmam
diagrams. Indeed, all events in the S picture used here are
related to the same instant $t=0$. Being aware of this, the line
directions in Fig.1 are given with the sole scope to discriminate
between nucleons and antinucleons. The latter will inevitably
appear in higher orders in coupling constants and for other
physical processes (e.g., the $\pi N$ scattering) as it has been
demonstrated in Ref. \cite{KorCanShe07}.

Further, for each boson included the corresponding relativistic
and properly symmetrized $N-N$ interaction, the kernel of integral
equations for the $N-N$ bound and scattering states, is determined
by
\begin{multline}
\left\langle b_c^\dag  (\vec p'_1) b_c^\dag (\vec p'_2) \Omega \right| K_b(NN \rightarrow NN) \left| b_c^\dag (\vec p_1) b_c^\dag (\vec p_2) \Omega \right\rangle \\
 = V_b^{dir}(1',2';1,2) - V_b^{exc}(1',2' ;1,2) \label{quasipot_definition}
\end{multline}
where we have separated the so--called direct
\begin{equation}
V_b^{dir} (1',2' ;1,2) =  - V_b (1',2' ;1,2) - V_b (2',1' ;2,1)
\end{equation}
and exchange
\begin{equation}
V_b^{exc} (1',2' ;1,2) = V_b^{dir} (2',1' ;1,2)
\end{equation}
terms.
For example, the one--pion--exchange contribution can be divided into the two parts:
\begin{multline}
V_{\pi}^{dir}(1',2';1,2) = - \frac{g_{\pi}^2}{(2\pi )^3} \frac{m^2}{\sqrt {E_{\vec p'_1} E_{\vec p'_2} E_{\vec p_1} E_{\vec p_2}}} \\
\times \delta \left( \vec p'_1  + \vec p'_2  - \vec p_1  - \vec p_2 \right) \bar u(\vec p'_1) \gamma_5 u(\vec p_1) \bar u(\vec p'_2) \gamma_5 u(\vec p_2) \\
\times \frac12 \left\{ \frac{1}{(p_1 - p'_1 )^2 - m_{\pi}^2} + \frac{1}{(p_2 - p'_2 )^2 - m_{\pi}^2} \right\}  \label{dir}
\end{multline}
and
\begin{multline}
V_{\pi}^{exc}(1',2';1,2) = - \frac{g_{\pi}^2}{(2\pi )^3} \frac{m^2}{\sqrt {E_{\vec p'_1} E_{\vec p'_2} E_{\vec p_1} E_{\vec p_2}}} \\
\times \delta \left( \vec p'_1  + \vec p'_2  - \vec p_1  - \vec p_2 \right) \bar u(\vec p'_1) \gamma_5 u(\vec p_2) \bar u(\vec p'_2) \gamma_5 u(\vec p_1) \\
\times \frac12 \left\{ \frac{1}{(p_2 - p'_1 )^2 - m_{\pi}^2} + \frac{1}{(p_1 - p'_2 )^2 - m_{\pi}^2} \right\}  \label{exc}
\end{multline}
to be depicted in Fig. 2, where the dashed lines correspond to the following Feynman--like "propagators":
$$
\frac12 \left\{ \frac{1}{(p_1 - p'_1 )^2 - m_{\pi}^2} + \frac{1}{(p_2 - p'_2 )^2 - m_{\pi}^2} \right\}
$$
on the left panel and
$$
\frac12 \left\{ \frac{1}{(p_2 - p'_1 )^2 - m_{\pi}^2} + \frac{1}{(p_1 - p'_2 )^2 - m_{\pi}^2} \right\}
$$
on the right panel.
Other distinctive features of the result (\ref{quasipot_definition}) have been discussed
in \cite{SheShi01,KorCanShe07}. Note also that expressions (\ref{dir})--(\ref{exc}) determine the
one--pion--exchange part of one--boson--exchange interaction
derived via the Okubo transformation method in \cite{KOSHE93} ( cf. \cite{FudaZhang95,KorCanShe07} )
taking into account the pion and heavier--meson exchanges.
\begin{figure}[h]
  \includegraphics[scale=0.55]{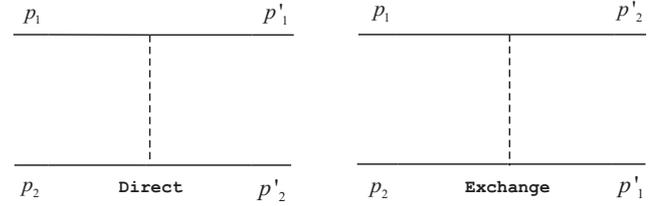}

\caption{The Feynman--like diagrams for the direct and exchange contributions in the r.h.s. of Eq.(23).}
\label{fig:2}
\end{figure}

\section{ The field--theoretic description of the elastic $N$--$N$ scattering }

\subsection{ The $T$--matrix in the CPR }

Usually in nonrelativistic quantum mechanics (NQM) the LS equation
for the T operator
\begin{equation}
T(E+i0)=V_{NN}+V_{NN}(E+i0-h_0)^{-1}T(E+i0)
\label{Toperator_definition}
\end{equation}
with a given kernel $V_{NN}$ is a starting point in evaluating the
$N$-$N$ phase shifts. All operators in Eq.
(\ref{Toperator_definition})  (including the sum $h_0$ of the
nucleon kinetic energies) act onto the subspace of two--nucleon
states, and remain confined in this subspace (the particle number
is conserved within nonrelativistic approach). The matrix elements
$
\langle N' N' | T(E+i0) | N N \rangle,
$
form the corresponding $T-$ matrix and, at the collision energy
$$
E=E_1 + E_2 = E_1^{'} + E_2^{'},
$$
the on--energy--shell elements can be expressed through the phase
shifts and mixing parameters (see below).

In relativistic QFT the situation is completely different. Though
one can formally introduce a field operator $T$ that meets the
equation
\begin{equation}
T(E+i0)=H_{I}+H_{I}(E+i0-H_F)^{-1}T(E+i0)
\end{equation}
the field interaction $H_I$, as a rule, does not conserve the
particle number, being the spring of particle creation and
destruction. The feature makes the problem of finding  the
$N$--$N$ scattering matrix much more complicated than in the framework of
nonrelativistic approach, since now the $T$ matrix enters
an infinite set of coupled integral equations.

Such a general field--theoretic consideration can be simplified with
the help of an equivalence theorem \cite{ShebekoFB17,SheISHEPP02} according to which the
$S$ matrix elements in the Dirac (\rm D) picture, viz.,
\begin{equation}
S_{fi} \equiv \langle \alpha^{\dag}...\Omega_0 | S(\alpha) |
\alpha^{\dag}...\Omega_0 \rangle
\label{Smatrix_bare}
\end{equation}
are equal to the corresponding elements
\begin{equation}
S_{fi}^c \equiv \langle \alpha_c^{\dag}...\Omega |
S(\alpha_c) | \alpha_c^{\dag}...\Omega \rangle
\label{Smatrix_clothed}
\end{equation}
of the $S$ matrix in the CPR. We say "corresponding" keeping in
mind the requirement $iii$) of Sec. 2. The $S$ operators in Eqs.
\eqref{Smatrix_bare}--\eqref{Smatrix_clothed} are determined by
the time evolution from the distant past to the distant future,
respectively, for the two decompositions
$$
H = H(\alpha) = H_F + H_I
$$
and
$$
H = K(\alpha_c) = K_F + K_I
$$
Note that the equality $S_{fi} = S_{fi}^c$ in question becomes
possible owing to certain isomorphism between the $\alpha_c$
algebra and the $\alpha$ algebra once the UCTs $W_D(t) =
\exp(iK_Ft)W\exp(-iK_Ft)$ obey the condition
\begin{equation}
W_D(\pm\infty)=1
\end{equation}
The $T$ operator in the CPR satisfies the equation
\begin{multline}
T_{cloth}(E+i0)=K_{I} \\
+K_{I}(E+i0-K_F)^{-1}T_{cloth}(E+i0)
\label{Toperator_clothed_definition}
\end{multline}
and the matrix
\begin{multline}
T_{fi} \equiv \langle f; b | T(E+i0) | i; b \rangle  \\
= \langle f; c| T_{cloth}(E+i0) | i; c \rangle \equiv T_{fi}^{c},
\end{multline}
where $| ; b\rangle$ ( $| ; c\rangle$ ) are the $H_F$ ( $K_F$ )
eigenvectors, may be evaluated relying upon properties of the new
interaction $K_I(\alpha_c)$. The latter has nonzero matrix
elements only between the clothed--particle (physical) states.
Such a restriction helps us to facilitate the further
consideration compared to the BPR with its large amount of virtual
transitions (cf., our discussion in Sect.1~).

If in Eq.(\ref{Toperator_clothed_definition}) we approximate
$K_I$ by $K_I^{(2)}$ (see Eq.(\ref{K(2)general_structure})),
then initial task of evaluating the BPR matrix elements $\langle
N' N' | T(E+i0) | N N \rangle$ can be reduced to solving the equation
\begin{multline}
\langle 1', 2' | T_{NN}(E) | 1, 2 \rangle = \langle 1', 2' | K_{NN} | 1, 2 \rangle \\
 +  \langle 1', 2' | K_{NN} (E+i0-K_F)^{-1} T_{NN}(E) | 1, 2 \rangle  \label{Tmatrix_equation}
\end{multline}
with $K_{NN} = K(NN \rightarrow NN)$.

Actually, let us employ the relation,
\begin{equation}
G_F(z) T_{cloth}(z) = G(z) K_I
\label{resolvente}
\end{equation}
with the two resolvents
$$
G_F(z)=(z-K_F)^{-1} \,\,\,\,\,\, G(z)=(z-K)^{-1}
$$
In its turn, Eq.(\ref{resolvente}) can be rewritten as
\begin{equation}
iG_F(z)T_{cloth}(z)= \int\limits_0^\infty d\tau e^{i(z-K)\tau} K_I
\end{equation}

To the approximation in question, where
\begin{equation}
K \simeq K_F+K_I^{(2)} \equiv K_2 + K_4 \label{K_aproximation}
\end{equation}
with
\begin{equation*}
K_2 = K_2(ferm) + K_2(mes),
\end{equation*}
\begin{equation*}
K_2(ferm) \sim b_c^{\dag} b_c + d_c^{\dag} d_c,\, K_2(mes) \sim a_c^{\dag}a_c
\end{equation*}
and
\begin{multline*}
K_4 \sim b_c^{\dag} b_c^{\dag} b_c b_c + d_c^{\dag} d_c^{\dag} d_c d_c + b_c^{\dag} d_c^{\dag} b_c d_c \\
+ a_c^{\dag} b_c^{\dag} a_c b_c + a_c^{\dag} d_c^{\dag} a_c d_c + d_c^{\dag} d_c^{\dag} a_c a_c + H.c.,
\end{multline*}
we have
\begin{equation*}
K_I b_c^{\dag} b_c^{\dag} | \Omega \rangle \simeq K_4 b_c^{\dag}
b_c^{\dag} | \Omega \rangle = K_4^{N} b_c^{\dag} b_c^{\dag} |
\Omega \rangle,
\end{equation*}
\begin{equation*}
K_4^{N} = b_c^{\dag} b_c^{\dag} b_c b_c + H.c.
\end{equation*}
so
\begin{equation*}
e^{i(z-K)\tau}K_I b_c^{\dag} b_c^{\dag} | \Omega \rangle \simeq
e^{iz\tau}e^{-i(K_2+K_4)\tau} K_4^{N} b_c^{\dag} b_c^{\dag}
|\Omega \rangle.
\end{equation*}
But
\begin{multline*}
e^{-i(K_2+K_4)\tau} K_4^{N} b_c^{\dag} b_c^{\dag} |\Omega \rangle = \left[ 1 - i(K_2+K_4)\tau  \phantom{\frac {i^{2}} {2!}} \right. \\
\left. + \frac {i^{2}} {2!} (K_2+K_4)^{2}\tau^{2} + \ldots \right] K_4^{N} b_c^{\dag} b_c^{\dag} |\Omega \rangle
\end{multline*}
with
$$
(K_2+K_4)K_4^{N} b_c^{\dag} b_c^{\dag} |\Omega \rangle =
(K_2^{N}+K_4^{N})K_4^{N} b_c^{\dag} b_c^{\dag} |\Omega \rangle,
$$
$$
(K_2+K_4)^{2}K_4^{N} b_c^{\dag} b_c^{\dag} |\Omega \rangle =
(K_2^{N}+K_4^{N})^{2}K_4^{N} b_c^{\dag} b_c |\Omega \rangle
$$
and so on. Here $ K_2^{N} \sim b_c^{\dag} b_c^{\dag} $.

Thus, to the approximation (\ref{K_aproximation})
\[
G_F(z)T(z)b_c^{\dag} b_c^{\dag} |\Omega \rangle
=(z - K_2^{N} - K_4^{N})^{-1}K_4^{N} b_c^{\dag} b_c^{\dag} |\Omega
\rangle
\]
or
\begin{equation}
T_{NN}(z)|1, 2\rangle = K_4^{N} |1, 2\rangle + K_4^{N}
(z-K_2^{N})^{-1} T_{NN}(z) | 1, 2 \rangle.
\label{Toperator_equation}
\end{equation}
Sometimes it is convenient to use the notation $ | 1, 2 \rangle =
b_c^{\dag} b_c^{\dag} |\Omega \rangle $ for any two nucleon state.
Equation (\ref{Tmatrix_equation}) follows from
(\ref{Toperator_equation}) if we take into account the
completeness condition
\begin{equation}
\int\limits_{NN} \!\!\!\!\!\!\!\!\! \sum |NN\rangle \langle NN| = 2
\end{equation}
and put $z = E+i0$.\footnote{Henceforth, the infinitesimal shift
$+i0$ will be omitted in $T(E+i0)$} Here the symbol $\int\limits_{NN} \!\!\!\!\!\!\!\! \sum$ means the
summation over nucleon polarizations and the integration over nucleon momenta.

\subsection{ The $R$--matrix equation and its angular--momentum decomposition.
The phase--shift relations }

For practical applications, in order to get rid of some discomfort
in handling the singularity of the resolvent $(E+i0-K_2^{N})^{-1}$, one
prefers to work with the $R$--matrix which is related to the
$T$--matrix by the Heitler equation ( see, e.g., Sect. 6 of
Chapter V in the monograph \cite{GW}):
\begin{equation}
T(E)=R(E)-i\pi R(E)\delta(E-K_2^{N})T(E) \label{Eq.41}
\end{equation}
Thus, for $R(E)$ in our case we obtain
\begin{multline}
R_{NN}(E)|1, 2 \rangle = K_{NN} |1, 2 \rangle \\
+ K_{NN} \frac{{\rm P}}{E-K_{2}^{N}} R_{NN}(E)|1, 2 \rangle,
\end{multline}
where ${\rm P}$ denotes the principal value ($p.v.$) to be
applied when the integration over the continuous spectrum of the
operator
\begin{equation*}
K_2^{N} = \sum\limits_{\mu} \int {d^3 pE_p b_c^\dag  \left( {\vec
p,\mu} \right) b_c \left( {\vec p,\mu} \right)}
\end{equation*}
is performed.

After this, one can write
\begin{multline}
\left\langle {1'2'} \right|\bar{R}(E)\left| {12} \right\rangle =
\left\langle {1'2'} \right|\bar{K}_{NN} \left| {12} \right\rangle \\
+ \int\limits_{34} \!\!\!\!\!\!\!\! \sum {\left\langle {1'2'} \right|\bar{K}_{NN}
\left| {34} \right\rangle \frac{{\left\langle {34}
\right|\bar{R}(E)\left| {12} \right\rangle }}{{E - E_3  - E_4 }}}
\label{R_int_equation1}
\end{multline}
with $\bar{R}(E) = R(E)/2$ and $\bar{K}_{NN} = K_{NN}/2 $, where
the operation $\int\limits_{34} \!\!\!\!\!\!\! \sum$ involves the $p.v.$ integration.

Certainly, the integral equation \eqref{R_int_equation1} has much in common with the
two--body $R$--matrix equation in NQM. It is true, unlike the
latter, in our case the center--of--mass motion is not separated
from the internal motion that is typical of relativistic theories
of interacting particles. The kernel of Eq.\eqref{R_int_equation1} is
\begin{multline*}
\left\langle {1'2'} \right|\bar K_{NN} \left| {12} \right\rangle =
\delta \left( \vec p'_1 + \vec p'_2 - \vec p_1  - \vec p_2  \right) \left\langle {1'2'} \right|\bar V\left| {12} \right\rangle \\
\equiv \delta \left( \vec p'_1 + \vec p'_2 - \vec p_1  - \vec p_2  \right) \\
\times \left\langle {\vec p'_1 \mu'_1 \tau'_1 , \vec p'_2 \mu'_2 \tau'_2 \left| {\bar V}
\right| \vec p_1 \mu _1 \tau _1 ,\vec p_2 \mu _2 \tau _2 } \right\rangle
\end{multline*}
that provides the total momentum conservation in every
intermediate state. The subsequent calculations are essentially simplified in
the center--of--mass system (c.m.s), in which we will employ the
notations
\begin{equation*}
\left| {\vec p \mu _1 \mu _2, \tau _1 \tau _2 }
\right\rangle = \left| {\vec p \mu _1 \tau _1 ,-\vec p \mu _2 \tau
_2 } \right\rangle,
\end{equation*}
\begin{equation*}
\left| {\vec{p}^{\,\prime} \mu _1' \mu _2',
\tau _1' \tau _2' } \right\rangle = \left| {\vec{p}^{\,\prime} \mu
_1' \tau _1' ,-\vec{p}^{\,\prime} \mu _2' \tau _2' }
\right\rangle
\end{equation*}
and
\begin{equation*}
\left| {\vec q \mu _3 \mu _4, \tau _3 \tau _4
} \right\rangle = \left| {\vec q \mu _3 \tau _3 ,-\vec q \mu _4
\tau _4 } \right\rangle,
\end{equation*}
respectively, for the initial, final and
intermediate states. Here the quantum numbers $\mu(\tau)$ are the
individual spin (isospin) projections.

Using these notations Eq.\eqref{R_int_equation1} in the c.m.s. can be written as
\begin{multline}
\left\langle \vec p' \mu'_1 \mu'_2, \tau'_1 \tau'_2 \left| {\bar R(E)} \right| \vec p \mu _1 \mu _2, \tau_1 \tau_2  \right\rangle  \\
\shoveright{= \left\langle {\vec p' \mu'_1 \mu'_2, \tau'_1 \tau'_2 \left| {\bar V} \right| \vec p \mu _1 \mu _2, \tau_1 \tau_2 } \right\rangle} \\
+ \sum {\rm P} \int d \vec q  \left\langle
\vec p' \mu'_1 \mu'_2, \tau'_1 \tau'_2 \left| {\bar V } \right| \vec q \mu _3 \mu _4, \tau_3 \tau_4 \right\rangle \\
\times \frac{\left\langle \vec q \mu _3 \mu _4, \tau_3 \tau_4 \left| {\bar R(E)} \right| \vec p \mu _1 \mu _2, \tau_1 \tau_2 \right\rangle}{E - 2E_{\vec q}}
\label{R_int_equation2}
\end{multline}
Accordingly Eq.\eqref{quasipot_definition}
\begin{multline}
\left\langle {1'2'} \right|{\bar V }\left| {12} \right\rangle =
-\frac{1}{2(2\pi )^3} \frac{m^2}{E_{\vec p'} E_{\vec p}} \\
\times \sum\limits_b [v_b^{dir} (1',2';1,2) - v_b^{exc}(1',2';2,1)]
\end{multline}
with
\begin{equation}
v_b^{dir} (1',2';1,2) = v_b(1',2';1,2) + v_b(2',1';2,1) \label{Eq.46}
\end{equation}
and
\begin{equation*}
v_b^{exc}(1',2';2,1) = v_b^{dir} (2',1';1,2),
\end{equation*}
where the separate boson contributions are determined by Eqs.
\eqref{Eq.20}--\eqref{Eq.22} with $\vec p_1 = \vec p = - \vec p_2$ and
$\vec p'_1 = \vec p' = -\vec p'_2$.

Following a common practice we are interested in the angular
--momentum decomposition of Eq.\eqref{R_int_equation2} assuming a nonrelativistic
analog of relativistic partial--wave expansions (see \cite{Werle66}
and refs. therein) for two--particle states. For example, the
clothed two--nucleon state (the so--called two--nucleon plane
wave) can be represented as
\begin{multline}
\left| \vec p \mu _1 \mu _2, \tau _1 \tau _2 \right\rangle  =
\sum \left( \frac12 \mu_1 \frac12 \mu_2 \left| SM_S \right. \right) \left( \frac12 \tau _1 \frac12 \tau _2 \left| TM_T \right. \right) \\
\left( lm_l SM_S \left| JM_J \right. \right) Y_{lm_l }^* \left( \vec p /p \right) \left| p J(lS)M_J ,TM_T \right\rangle,
\label{Eq.47}
\end{multline}
where, as shown in Appendix B, $J$, $S$ and $T$ are, respectively,
total angular momentum, spin and isospin of the $NN$ pair. All
necessary summations over dummy quantum numbers are implied.

Of course, when deriving expansions like \eqref{Eq.47} (see Appendix B) one
needs to take into account a distinctive feature of any QFT
associated with particle production and annihilation. The use of
such expansions gives rise to the well known JST representation
(see, e.g., \cite{GW}), in which
\begin{multline}
\left\langle p' J'(l'S')M'_J, T'M'_T \left| {\bar V} \right| p J(lS)M_J, TM_T  \right\rangle \\
= \bar V_{l'l}^{JST} (p',p) \delta _{J'J} \delta _{M'_J M_J } \delta _{S'S} \delta _{T'T} \delta _{M'_T M_T }, \label{Eq.48}
\end{multline}
\begin{multline}
\left\langle p' J'(l'S')M'_J, T'M'_T \left| {\bar R (E)} \right| p J(lS)M_J, TM_T \right\rangle \\
= \bar R_{l'l}^{JST} (p',p;E)\delta _{J'J} \delta _{M'_J M_J } \delta _{S'S} \delta _{T'T} \delta _{M'_T M_T }, \label{Eq.49}
\end{multline}
so Eq.\eqref{R_int_equation2} reduces to the set of simple integral equations,
\begin{multline}
\bar R_{l'\,l}^{JST}(p',p) = \bar V_{l'\,l}^{JST}(p',p) \\
+ \sum\limits_{l''} {\rm P} \int\limits_0^\infty
\frac{q^2 \, dq}{2(E_p  - E_q )} \bar V_{l'\,l''}^{JST}(p',q) \bar R_{l''{\rm{ }}l}^{JST}(q,p)
\label{Eq.50}
\end{multline}
to be solved for each submatrix $\overline{R}^{JST}$ composed of
the elements
\begin{equation}
\bar R_{l'l}^{JST}(p',p) \equiv
\bar R_{l'l}^{JST}(p',p;2E_p), \label{Eq.51}
\end{equation}
where $2E_p = 2\sqrt{\vec p^2 + m^2}$ represents the collision
energy in the c.m.s. As usually, we distinguish the
on--energy--shell ($p'=p$) and half--off--energy--shell ($p' \neq
p$) elements. Apparently, it is pertinent to stress that even if
we are interested in the $R-$matrix on the energy shell one has
inevitably to go out beyond it.

In addition, one should note that in view of the charge independence assumed in
this work one has to solve two separate equations for isospin
values $T=0$ and $T=1$. Regarding the $T-$ dependence of the
corresponding solutions in details one needs to keep in mind the
typical factor $[1 - (-1)^{l+S+T}]$ that is discussed in Appendix C
(below Eq.(C.24)). It means such a selection of the partial $R-$
equations, where the number $l+S+T$ must be odd.

Properties \eqref{Eq.48} and \eqref{Eq.49} are resulted from rotational invariance,
parity and isospin conservation in combination with the
antisymmetry requirement for two--nucleon states. At the point,
let us recall that the clothing procedure violates no one of these
symmetries (details can be found in survey \cite{SheShi01}). In
particular, it implies that
\begin{multline}
\left[ {\vec J_F (\alpha _c ),K_I (\alpha _c )} \right] = \left[
{\vec J_F (\alpha _c ),K_I^{(2)} (\alpha _c )} \right] \\
= \left[{\vec J_{ferm} ,K(NN \to NN)} \right] = 0,
\end{multline}
\begin{multline}
\left[ {\mathscr{P}(\alpha _c ),K_I (\alpha _c )} \right] = \left[
{\mathscr{P}(\alpha _c ),K_I^{(2)} (\alpha _c )} \right] \\
= \left[ {\mathscr{P}_{ferm} ,K(NN \to NN)} \right] = 0
\end{multline}
and
\begin{equation}
\left[ {\vec J_F (\alpha _c ),S(\alpha _c )} \right] = \left[
{\mathscr{P}(\alpha _c ),S(\alpha _c )} \right] = 0,
\end{equation}
where $\vec{J}_F(\alpha_c)=\vec{J}$ the operator of total angular
momentum, $\mathscr{P}(\alpha_c)=\mathscr{P}$ the operator of
space inversion. As in Appendix B, we distinguish their fermionic
parts $\vec{J}_{ferm}$ and $\mathscr{P}_{ferm}$ determined by Eqs.
\eqref{(B.6)} and \eqref{(B.21)}.

Further, apart from the transition with $l=l'=J$ and $l(l')=J \pm
1$ all different ones are forbidden owing to the parity
conservation. Therefore, for each $JST-$ channel there are only
six independent elements~\footnote{Below in this Sect., for
brevity, we shall write $R(V)$ instead of $\bar R(\bar V)$.}:
\begin{equation*}
R_{JJ}^{JS=0T}(p',p), \,\, R_{JJ}^{JS=1T}(p',p), \,\, R_{J-1\,J-1}^{JS=1T}(p',p),
\end{equation*}
\begin{equation*}
R_{J-1\,J+1}^{JS=1T}(p',p), \,\, R_{J+1\,J-1}^{JS=1T}(p',p), \,\, R_{J+1\,J+1}^{JS=1T}(p',p).
\end{equation*}
The first two of them satisfy the equations:

for spin singlet
\begin{multline}
R_{JJ}^{JS=0T}(p',p) = V_{JJ}^{JS=0T}(p',p) \\
+\frac{1}{2} {\rm P}
\int \limits_0^{\infty} \frac{q^2\,dq}{E_p -E_q}
V_{JJ}^{JS=0T}(p',q)R_{JJ}^{JS=0T}(q,p) , \label{Eq.55}
\end{multline}
and for uncoupled spin triplet
\begin{multline}
R_{JJ}^{JS=1T}(p',p) = V_{JJ}^{JS=1T}(p',p) \\
+\frac{1}{2} {\rm P}
\int \limits_0^{\infty} \frac{q^2\,dq}{E_p -E_q}
V_{JJ}^{JS=1T}(p',q)R_{JJ}^{JS=1T}(q,p) ,
\end{multline}
while for finding the rest one needs to solve the set of four
coupled integral equations
\begin{multline*}
R_{J - 1\,J - 1}^{JS = 1T} (p',p) = V_{J - 1\,J - 1}^{JS = 1T} (p',p) \\
+ \frac{{\rm P}}{2}\int \frac{q^2 \,dq}{E_p  - E_q} \left[ V_{J - 1\,J - 1}^{JS = 1T}(p',p)R_{J - 1\,J - 1}^{JS = 1T}(p',p) \right. \\
\left. + V_{J - 1\,J + 1}^{JS = 1T}(p',p)R_{J + 1\,J - 1}^{JS = 1T}(p',p) \right]
\end{multline*}
\begin{multline*}
R_{J + 1\,J - 1}^{JS = 1T}(p',p) = V_{J + 1\,J - 1}^{JS = 1T} (p',p) \\
+ \frac{{\rm P}}{2}\int \frac{q^2 \,dq}{E_p  - E_q}\left[ V_{J + 1\,J - 1}^{JS = 1T}(p',p)R_{J - 1\,J - 1}^{JS = 1T}(p',p) \right. \\
\left. + V_{J + 1\,J + 1}^{JS = 1T}(p',p)R_{J + 1\,J - 1}^{JS = 1T} (p',p) \right]
\end{multline*}
\begin{multline*}
R_{J - 1\,J + 1}^{JS = 1T} (p',p) = V_{J - 1\,J + 1}^{JS = 1T} (p',p) \\
+ \frac{{\rm P}}{2}\int \frac{q^2 \,dq}{E_p  - E_q} \left[ V_{J - 1\,J - 1}^{JS = 1T}(p',p)R_{J - 1\,J + 1}^{JS = 1T}(p',p) \right. \\
\left. + V_{J - 1\,J + 1}^{JS = 1T}(p',p)R_{J + 1\,J + 1}^{JS = 1T}(p',p) \right]
\end{multline*}
\begin{multline}
R_{J + 1\,J + 1}^{JS = 1T}(p',p) = V_{J + 1\,J + 1}^{JS = 1T} (p',p) \\
+ \frac{{\rm P}}{2}\int \frac{q^2 \,dq}{E_p  - E_q}\left[ V_{J + 1\,J - 1}^{JS = 1T}(p',p)R_{J - 1\,J + 1}^{JS = 1T}(p',p) \right. \\
\left. + V_{J + 1\,J + 1}^{JS = 1T}(p',p)R_{J + 1\,J + 1}^{JS = 1T} (p',p) \right]
\label{Eq.57}
\end{multline}
One should stress that after omitting the factor
$[1 - (-1)^{l+S+T}]$ the matrix elements $V_{l'l}^{JST}$ in these
equations are given by the direct parts of our energy independent
quasipotentials, which are derived in Appendix C.

The Heitler equation \eqref{Eq.41} enables one to get the partial $T-$
matrix elements
\begin{multline}
T_{l'l}^{JST}(p',p) = R_{l'l}^{JST}(p',p) \\
-i\pi \rho(p) \sum \limits_{l''} R_{l'l''}^{JST}(p',p) T_{l''l}^{JST}(p,p),
\end{multline}
whence it follows the on--shell relationship:
\begin{equation}
T_{l'l}^{JST}(p) = R_{l'l}^{JST}(p) -i\pi \rho(p) \sum
\limits_{l''} R_{l'l''}^{JST}(p) T_{l''l}^{JST}(p),
\end{equation}
where
$$
T(R)_{l'l}^{JST}(p) \equiv T(R)_{l'l}^{JST}(p,p;E=2E_p)
$$
and $\rho(p) = pE_p/2$.

In turn, with the help of a standard definition of the on--shell
$S$--matrix elements
\begin{equation}
S_{JJ}^{JST}(p) \equiv \exp \left[ 2i \delta_J^{ST}\right] = 1 - 2
\pi i \rho(p) T_{JJ}^{JST}(p)
\end{equation}
the $R$--matrix elements for the uncoupled states can be expressed
through the phase shifts $\delta_J^{ST}$:
\begin{equation}
- \pi \rho(p) R_{JJ}^{JST}(p) = \tan \delta_J^{ST}
\end{equation}
Usually the isospin label is suppressed to write simply
$\delta_J^S$. Of course, these quantities depend either on the
incoming momentum $p$ or the laboratory energy $E_{lab}$. We
choose the second alternative.

For the coupled states the on--shell $R-$ matrix elements are
conventionally parameterized in terms of the phase shifts
$\delta_{\pm}^J$ and the mixing parameters $\varepsilon_J$:
\begin{multline}
 - \pi \rho (p)\left( {\begin{array}{*{20}c}
   {R_{J - 1\,J - 1}^{J1} } & {R_{J - 1\,J + 1}^{J1} }  \\
   {R_{J + 1\,J - 1}^{J1} } & {R_{J + 1\,J + 1}^{J1} }  \\
\end{array}} \right)
 = \left( {\begin{array}{*{20}c}
   {\cos \varepsilon _J } & { - \sin \varepsilon _J }  \\
   {\sin \varepsilon _J } & {\cos \varepsilon _J }  \\
\end{array}} \right) \\
\times \left( {\begin{array}{*{20}c}
   {\exp [2i\delta _ - ^J]} & 0  \\
   0 & {\exp [2i\delta _ + ^J]}  \\
\end{array}} \right)\left( {\begin{array}{*{20}c}
   {\cos \varepsilon _J } & {\sin \varepsilon _J }  \\
   { - \sin \varepsilon _J } & {\cos \varepsilon _J }  \\
\end{array}} \right)
\label{Eq.62}
\end{multline}
Such a parametrization was put forward in \cite{BlBied}. From
Eq.\eqref{Eq.62} it follows that
\begin{multline}
\tan \delta_{\pm}^J = -\frac{1}{2}\pi \rho(p) \left[
R_{J+1\,J+1}^{J1} + R_{J-1\,J-1}^{J1} \phantom{\frac{R_{J+1\,J-1}^{J1}
+ R_{J-1\,J+1}^{J1}}{\cos \varepsilon_J}} \right. \\
\left. \mp \frac{R_{J+1\,J-1}^{J1}
+ R_{J-1\,J+1}^{J1}}{\cos \varepsilon_J} \right]
\end{multline}
and
\begin{equation}
\tan 2\varepsilon_J = \frac{R_{J+1\,J-1}^{J1} +
R_{J-1\,J+1}^{J1}}{R_{J-1\,J-1}^{J1} - R_{J+1\,J+1}^{J1}}
\end{equation}
Here in $R_{l'l}^{J1T}(p)$ the isospin index and the argument $p$
are suppressed too.

However, in the next section our calculations are shown for the
so--called bar convention introduced in \cite{Stapp}. These are
related to the Blatt--Biedenharn phase shifts by
\begin{equation*}
\bar \delta_{+}^J + \bar \delta_{-}^J = \delta_{+}^J + \delta_{-}^J, \,\,\,\,
\sin(\bar \delta_{-}^J - \bar \delta_{+}^J) = \frac{\tan 2\bar \varepsilon_J}{\tan 2\varepsilon_J},
\end{equation*}
\begin{equation}
\sin(\delta_{-}^J - \delta_{+}^J) = \frac{\sin 2\bar \varepsilon_J}{\sin 2\varepsilon_J}
\end{equation}
(cf. Eq.\eqref{(C.17)} from \cite{MachHolElst87}). It will allow us to
compare our results directly with those by the Bonn group (in
particular, from the survey \cite{Mach89}).

\section{Results of numerical calculations and their discussion}

Available experience of solving integral equations similar to Eqs.
\eqref{Eq.55}--\eqref{Eq.57} shows that it is convenient to employ
the so--called matrix inversion method (MIM)
\cite{Brown_et_al_69}, \cite{HaftTabakin70} (more sophisticated
methods are discussed in the monograph \cite{BJ76}). In the course
of our numerical calculations we have improved a code
\cite{KorchinShebeko77} based upon the MIM and successfully
applied for the treatment of the final--state interaction in
studies \cite{KorchinShebeko90}, \cite{LadyShe04} of the deuteron
breakup by electrons and protons in the GeV region.

Since we deal with the relativistic dispersion law for the particle
energies, the well known substraction procedure within the MIM
in our case leads to equations
\begin{multline}
R_{l'{\rm{ }}l}^{JST} (p',p) = V_{l'{\rm{ }}l}^{JST}(p',p) \\
+ \frac12 \sum \limits_{l''} \int \limits_0^\infty \frac{d q}{p^2 - q^2}
\left\{ q^2 (E_p  + E_q) V_{l'l''}^{JST}(p',q) R_{l''l}^{JST}(q,p) \right. \\
\left. - 2p^2 E_p V_{l'l''}^{JST}(p',p) R_{l''l}^{JST}(p,p) \right\} . \label{Eq.66}
\end{multline}
To facilitate comparison with some derivations and calculations
from Refs. \cite{MachHolElst87}, \cite{Mach89}, we introduce the
notation
\begin{multline*}
\left\langle {\vec p'\,\mu'_1 \mu' _2 } \right| v^{UCT}_b \left|
{\vec p\,\mu _1 \mu _2 } \right\rangle \\ \equiv
-F^2_b(p',p)
\left[ v_b (1',2';1,2)+v_b (2',1' ;2,1) \right]
\end{multline*}
for the regularized UCT quasipotentials
in the c.m.s. (see Appendix C). As in Ref.\cite{MachHolElst87}, we put
\begin{equation*}
F_b(p',p) = \left[ \frac{\Lambda _b^2  - m_{b}^2}{\Lambda _{b}^2
- (p' - p)^2} \right]^{n_{b}} \equiv F_b[(p'-p)^2]
\end{equation*}
Doing so, we have
\begin{multline}
\left\langle \vec p^{\, \prime} \,\mu'_1 \mu' _2  \right|   v^{UCT}_{s} \left| \vec p \,\mu _1 \mu _2 \right\rangle \\
= g_s^2 \bar u( \vec p^{\, \prime}) u(\vec p ) \frac{F^2_{s}[(p'-p)^2]}{(p' - p )^2  - m_s^2} \bar u( - \vec p^{\, \prime} ) u( - \vec p ),
\label{Eq.67}
\end{multline}
\begin{multline}
\left\langle \vec p^{\, \prime} \,\mu'_1 \mu' _2 \right|   v^{UCT}_{ps} \left| \vec p\,\mu _1 \mu _2 \right\rangle \\
= - g_{ps}^2 \bar u(\vec p^{\, \prime}) \gamma _5 u(\vec p)  \frac{F^2_{ps}[(p'-p)^2]}{(p' - p )^2 - m_{ps}^2} \bar u( - \vec p^{\, \prime}) \gamma _5 u( - \vec p )
\label{Eq.68}
\end{multline}
and
\begin{multline}
\left\langle {\vec p^{\, \prime} \,\mu'_1 \mu' _2 } \right|   v^{UCT}_{\rm {v}} \left| {\vec p\,\mu _1 \mu _2 } \right\rangle
= - \frac{F^2_{\rm v}[(p'-p)^2]}{\left({p' - p}\right)^2  - m_{\rm{v}}^2 } \\
\shoveleft{ \times \left\{ \bar u (\vec p^{\, \prime}) \left[ \left( g_{\rm{v}} + f_{\rm{v}} \right) \gamma _{\nu} - \frac{f_{\rm{v}}}{2m} \left( p' + p \right)_{\nu } \right. \right.}\\
\shoveright{ \left. - \frac{f_{\rm{v}}}{2m} (E_{\vec p'} - E_{\vec p})[\gamma_0 \gamma_{\nu}-g_{0 \nu}] \right] u\left(\vec p\right) }  \\
\shoveleft{\times \bar u\left(- \vec p^{\, \prime} \right) \left[ \left({g_{\rm{v}} + f_{\rm{v}}} \right) \gamma ^\nu - \frac{f_{\rm{v}}}{2m} \overline{\left( p' + p \right)}^{\nu} \right.}  \\
\shoveright{ \left. - \frac{f_{\rm{v}}}{2m} (E_{\vec p'} - E_{\vec p})[\gamma^0 \gamma^{\nu}-g^{0 \nu}] \right]u(- \vec p) } \\
\shoveleft{- \frac{{f_{\rm v}}^2}{4m^2} (E_{p'}-E_p)^2 \bar u(\vec p^{\, \prime})[\gamma_0 \gamma_{\nu}-g_{0 \nu}]u(\vec p)} \\
\left. \phantom{\frac{f_{\rm{v}}}{2m}} \times \bar u(-\vec p^{\,
\prime})[\gamma^0 \gamma^{\nu}-g^{0 \nu}]u(-\vec p) \right\},
\label{Eq.69}
\end{multline}
where $\overline{( p' + p)}^{\nu} = (E_{\vec p'} + E_{\vec p},
-(\vec p' + \vec p))$.

\begin{table}[!ht]
\caption{The best--fit parameters for the two models. The third (fourth) column taken from Table A.1 \cite{Mach89}
(obtained by solving Eqs.\eqref{Eq.66} with a least squares fitting \cite{Holland,Kor05} to OBEP values in Table 2). All masses
are in $MeV$, and $n_b=1$ except for $n_{\rho}=n_{\omega}=2.$}
\label{UCT_tab:1} \tabcolsep=0.05\columnwidth

\begin{tabular}{cccc}

\hline \hline \noalign{\smallskip}

Meson &  & Potential B & UCT \\

\noalign{\smallskip}\hline\noalign{\smallskip}

$\pi$    & $g^2_{\pi}/4\pi$  & 14.4   & 14.5    \\
         & $\Lambda_{\pi}$   & 1700   & 2200    \\
         & $m_{\pi}$         & 138.03 & 138.03  \\

\noalign{\smallskip}\hline\noalign{\smallskip}

$\eta$    & $g^2_{\eta}/4\pi$  & 3     & 2.8534  \\
          & $\Lambda_{\eta}$   & 1500  & 1200    \\
          & $m_{\eta}$         & 548.8 & 548.8   \\

\noalign{\smallskip}\hline\noalign{\smallskip}

$\rho$    & $g^2_{\rho}/4\pi$    & 0.9  & 1.3   \\
          & $\Lambda_{\rho}$     & 1850 & 1450  \\
          & $f_{\rho}/g_{\rho}$  & 6.1  & 5.85  \\
          & $m_{\rho}$           & 769  & 769   \\

\noalign{\smallskip}\hline\noalign{\smallskip}

$\omega$    & $g^2_{\omega}/4\pi$  & 24.5  & 27      \\
            & $\Lambda_{\omega}$   & 1850  & 2035.59 \\
            & $m_{\omega}$         & 782.6 & 782.6   \\

\noalign{\smallskip}\hline\noalign{\smallskip}

$\delta$    & $g^2_{\delta}/4\pi$  & 2.488 & 1.6947  \\
            & $\Lambda_{\delta}$   & 2000  & 2200    \\
            & $m_{\delta}$         & 983   & 983     \\

\noalign{\smallskip}\hline\noalign{\smallskip}

$\sigma, \,T=0 $    & $g^2_{\sigma}/4\pi$  & 18.3773   & 19.4434   \\
                    & $\Lambda_{\sigma}$   & 2000      & 1538.13   \\
                    & $m_{\sigma}$         & 720       & 717.72  \\

\noalign{\smallskip}\hline\noalign{\smallskip}

$\sigma, \, T=1 $    & $g^2_{\sigma}/4\pi$  & 8.9437    &  10.8292   \\
                     & $\Lambda_{\sigma}$   & 1900      &  2200      \\
                     & $m_{\sigma}$         & 550       &  568.86  \\

\noalign{\smallskip}\hline\noalign{\smallskip}

\end{tabular}
\end{table}

At first sight, such a regularization can be achieved via a simple substitution $g_b \rightarrow g_b F_b(p',p)$ with some cutoff functions $F_b(p',p)$ depending on the 4--momenta $p'$ and~$p$.
\begin{table*}[!ht]
\caption{Neutron--proton phase shifts (in degrees) for various
laboratory energies (in MeV). The OBEP(OBEP$^*$)--rows taken from Table~5.2 \cite{Mach89} (calculated by solving Eqs.\eqref{Eq.71} with the model parameters from the third column in Table 1). The UCT$^*$(UCT)--rows calculated by solving Eqs.\eqref{Eq.66} with the parameters from the third (fourth) column in Table 1. As in \cite{MachHolElst87}, we have used the bar convention \cite{Stapp} for the phase parameters.}
\label{UCT_tab:2} \tabcolsep=0.08\columnwidth
\begin{center}
\begin{tabular}{cccccccc}

\hline \hline \noalign{\smallskip}

State & Potential & 25 & 50 & 100 & 150 & 200 & 300 \\

\noalign{\smallskip}\hline\noalign{\smallskip}

            & OBEP     & 50.72 & 39.98 & 25.19 & 14.38 & 5.66 & -8.18  \\
{${^1}S_0$} & OBEP$^*$ & 50.71 & 39.98 & 25.19 & 14.37 & 5.66 & -8.18  \\
            & UCT$^*$  & 66.79 & 53.01 & 36.50 & 25.27 & 16.54 & 3.12  \\
            & UCT      & 50.03 & 39.77 & 25.55 & 15.20 & 6.92 & -6.07  \\

\noalign{\smallskip}\hline\noalign{\smallskip}

            & OBEP     & -7.21 & -11.15 & -16.31 & -20.21 & -23.47 & -28.70  \\
{${^1}P_1$} & OBEP$^*$ & -7.17 & -11.15 & -16.32 & -20.21 & -23.48 & -28.71  \\
            & UCT$^*$  & -7.40 & -11.70 & -17.73 & -22.63 & -26.98 & -34.54  \\
            & UCT      & -7.15 & -10.95 & -15.62 & -18.90 & -21.49 & -25.41  \\

\noalign{\smallskip}\hline\noalign{\smallskip}

            & OBEP     & 0.68 & 1.58 & 3.34 & 4.94 & 6.21 & 7.49  \\
{${^1}D_2$} & OBEP$^*$ & 0.68 & 1.58 & 3.34 & 4.94 & 6.21 & 7.49  \\
            & UCT$^*$  & 0.68 & 1.59 & 3.40 & 5.10 & 6.52 & 8.20  \\
            & UCT      & 0.68 & 1.56 & 3.22 & 4.68 & 5.77 & 6.68  \\

\noalign{\smallskip}\hline\noalign{\smallskip}

            & OBEP     & 9.34 & 12.24 & 9.80  & 4.57 & -1.02 & -11.48   \\
{${^3}P_0$} & OBEP$^*$ & 9.34 & 12.24 & 9.80  & 4.57 & -1.02 & -11.48   \\
            & UCT$^*$  & 9.48 & 12.53 & 10.32 & 5.27 & -0.15 & -10.27   \\
            & UCT      & 9.30 & 12.16 & 9.81  & 4.73 & -0.68 & -10.76   \\

\noalign{\smallskip}\hline\noalign{\smallskip}

            & OBEP     & -5.33 & -8.77 & -13.47 & -17.18 & -20.49 & -26.38  \\
{${^3}P_1$} & OBEP$^*$ & -5.33 & -8.77 & -13.47 & -17.18 & -20.48 & -26.38  \\
            & UCT$^*$  & -5.27 & -8.62 & -13.09 & -16.56 & -19.63 & -25.06  \\
            & UCT      & -5.28 & -8.58 & -12.85 & -16.06 & -18.86 & -23.79  \\

\noalign{\smallskip}\hline\noalign{\smallskip}

            & OBEP     & 3.88 & 9.29 & 17.67 & 22.57 & 24.94 & 25.36  \\
{${^3}D_2$} & OBEP$^*$ & 3.89 & 9.29 & 17.67 & 22.57 & 24.94 & 25.36  \\
            & UCT$^*$  & 3.86 & 9.15 & 17.12 & 21.51 & 23.47 & 23.48  \\
            & UCT      & 3.89 & 9.25 & 17.31 & 21.77 & 23.75 & 23.61  \\

\noalign{\smallskip}\hline\noalign{\smallskip}

            & OBEP     & 80.32 & 62.16 & 41.99 & 28.94 & 19.04 & 4.07   \\
{${^3}S_1$} & OBEP$^*$ & 80.31 & 62.15 & 41.98 & 28.93 & 19.03 & 4.06   \\
            & UCT$^*$  & 92.30 & 72.71 & 51.44 & 38.10 & 28.20 & 13.70  \\
            & UCT      & 79.60 & 61.53 & 41.57 & 28.75 & 19.08 & 4.60  \\

\noalign{\smallskip}\hline\noalign{\smallskip}

            & OBEP     & -2.99 & -6.86 & -12.98 & -17.28 & -20.28 & -23.72  \\
{${^3}D_1$} & OBEP$^*$ & -2.99 & -6.87 & -12.99 & -17.28 & -20.29 & -23.72  \\
            & UCT$^*$  & -2.74 & -6.43 & -12.36 & -16.54 & -19.47 & -22.78  \\
            & UCT      & -3.00 & -6.90 & -13.12 & -17.66 & -21.11 & -26.03  \\

\noalign{\smallskip}\hline\noalign{\smallskip}

                  & OBEP     & 1.76  & 2.00  & 2.24  & 2.58  & 3.03 & 4.03  \\
{$\varepsilon_1$} & OBEP$^*$ & 1.76  & 2.00  & 2.24  & 2.58  & 3.03 & 4.03  \\
                  & UCT$^*$  & 0.02  & -0.12 & -0.17 & 0.04  & 0.41 & 1.40  \\
                  & UCT      & 1.80  & 2.01  & 2.19  & 2.50  & 2.90 & 3.83  \\

\noalign{\smallskip}\hline\noalign{\smallskip}

            & OBEP    & 2.62 & 6.14 & 11.73 & 14.99 & 16.65 & 17.40  \\
{${^3}P_2$} & OBEP$^*$& 2.62 & 6.14 & 11.73 & 14.99 & 16.65 & 17.39  \\
            & UCT$^*$ & 2.80 & 6.61 & 12.71 & 16.28 & 18.10 & 18.91  \\
            & UCT     & 2.57 & 6.00 & 11.32 & 14.18 & 15.37 & 15.07  \\

\noalign{\smallskip}\hline\noalign{\smallskip}

            & OBEP     & 0.11 & 0.34 & 0.77 & 1.04 & 1.10 & 0.52  \\
{${^3}F_2$} & OBEP$^*$ & 0.11 & 0.34 & 0.77 & 1.04 & 1.10 & 0.52  \\
            & UCT$^*$  & 0.11 & 0.34 & 0.77 & 1.05 & 1.13 & 0.64  \\
            & UCT      & 0.11 & 0.34 & 0.75 & 1.00 & 1.03 & 0.41  \\

\noalign{\smallskip}\hline\noalign{\smallskip}

                  & OBEP     & -0.86 & -1.82 & -2.84 & -3.05 & -2.85 & -2.02  \\
{$\varepsilon_2$} & OBEP$^*$ & -0.86 & -1.82 & -2.84 & -3.05 & -2.85 & -2.02  \\
                  & UCT$^*$  & -0.87 & -1.83 & -2.82 & -2.99 & -2.75 & -1.88  \\
                  & UCT      & -0.86 & -1.83 & -2.84 & -3.05 & -2.89 & -2.18  \\

\noalign{\smallskip}\hline\noalign{\smallskip}

\end{tabular}
\end{center}
\end{table*}
\begin{figure*}[!ht]
\begin{center}
  \includegraphics[scale=0.80]{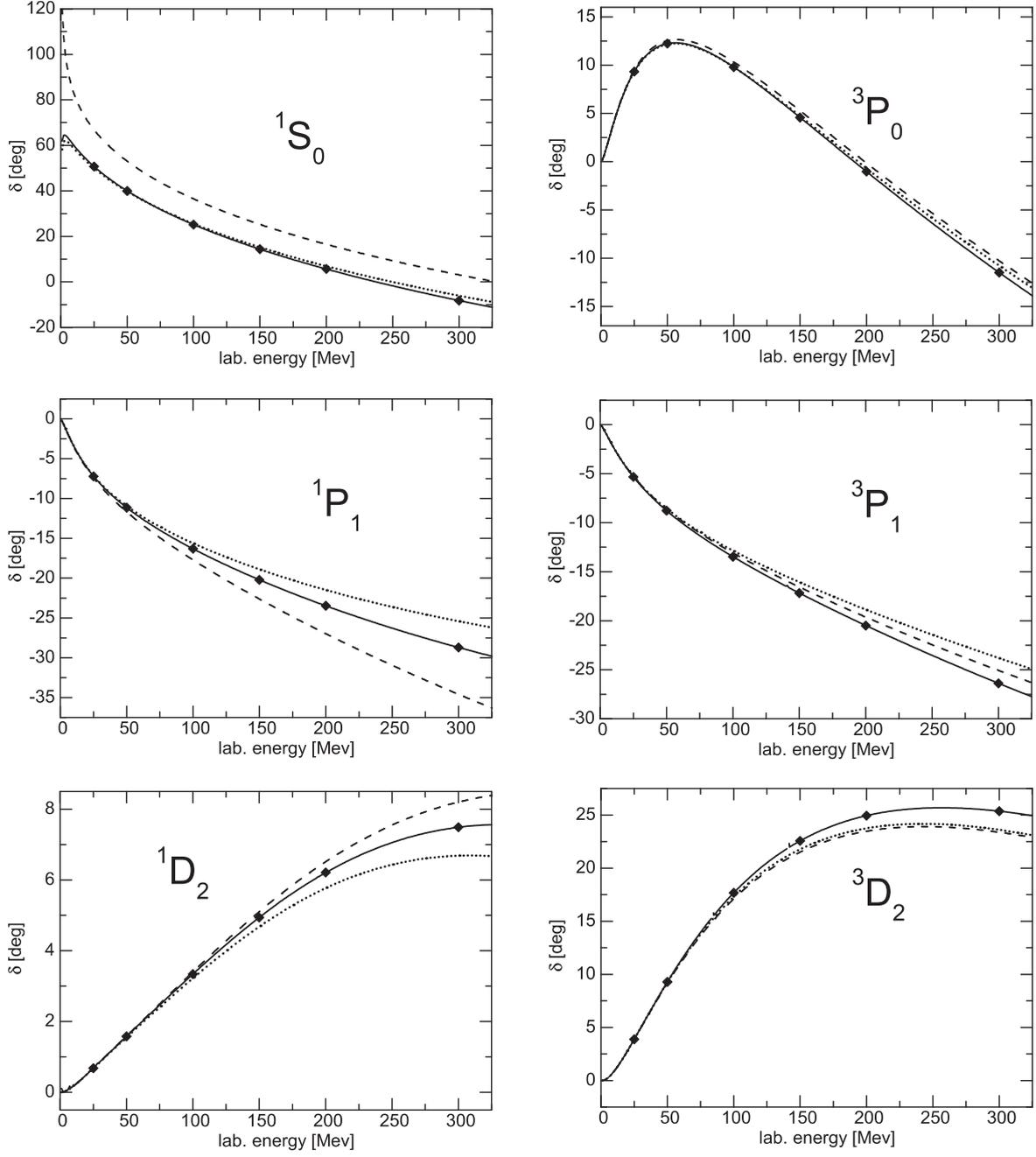}
\end{center}
\caption{Neutron-proton phase  parameters for the uncoupled
partial waves, plotted versus the nucleon kinetic energy in the lab. system.
Dashed[solid] curves calculated with Potential B parameters (Table \ref{UCT_tab:1})  by solving Eqs. \eqref{Eq.66}[\eqref{Eq.71}]. Dotted represent the solutions of Eqs. \eqref{Eq.66} with UCT parameters (Table \ref{UCT_tab:1}). The rhombs show original OBEP results (see Table 2).}
\label{UCT_fig:3}
\end{figure*}

However, a principal moment is to satisfy the requirement (86) for the Hamiltonian invariant under space inversion, time reversal and charge conjugation. A constructive consideration of the issue is given in Appendix~ C.
{\sloppy

}
Replacing in equations \eqref{Eq.67}--\eqref{Eq.69}
\begin{equation*}
F^2_b[(p'-p)^2] \left\{ (p' - p)^2  - m_b^2 \right\}^{-1}
\end{equation*}
by
\begin{equation*}
-F^2_b[-(\vec p'- \vec p)^2] \left\{ (\vec p' - \vec p)^2 + m_b^2 \right\}^{-1}
\end{equation*}
and neglecting the tensor-tensor term
\begin{figure*}[t]
\begin{center}
  \includegraphics[scale=0.80]{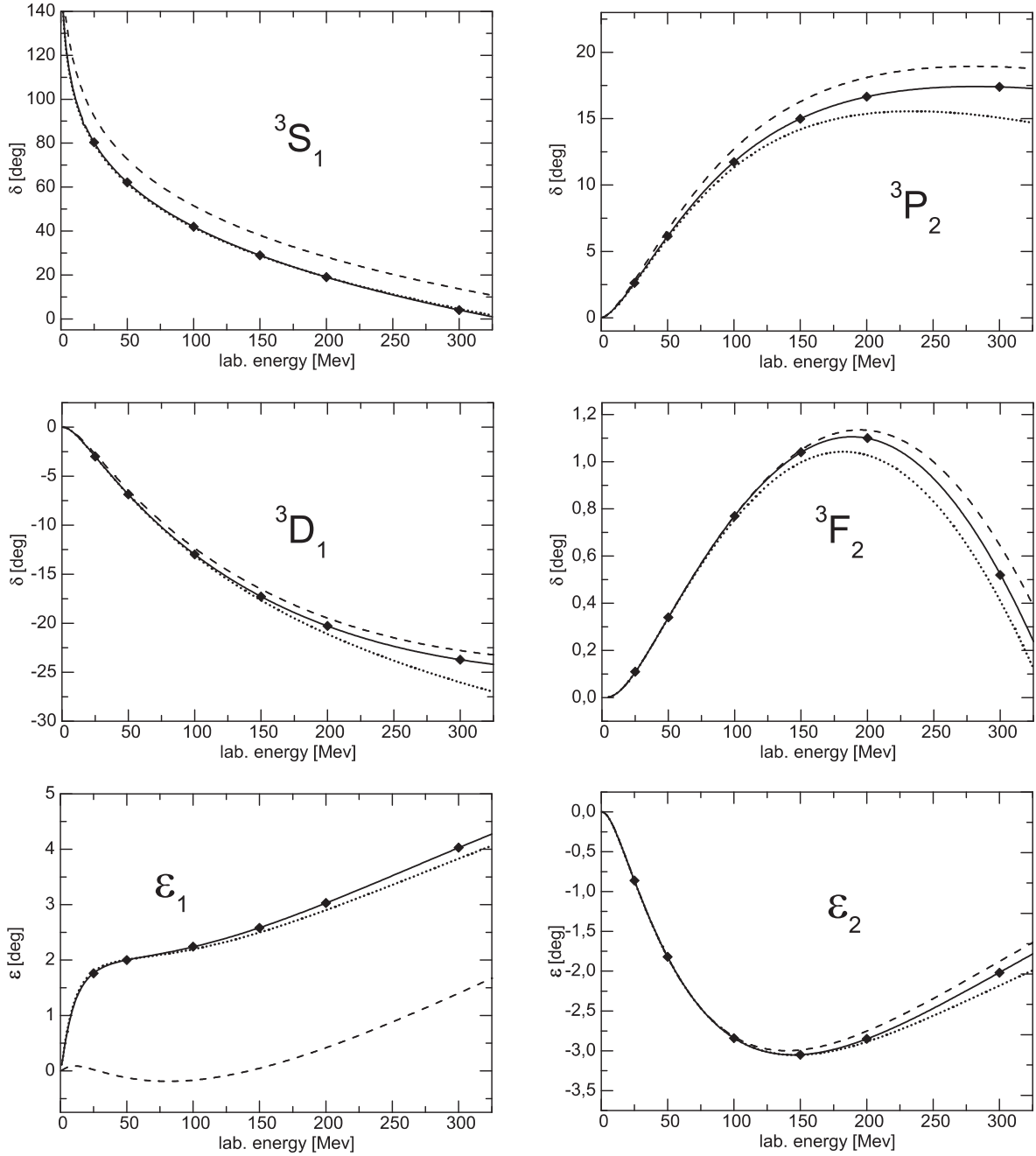}
\end{center}

\caption{The same in Fig.3 but for the coupled waves.}
\label{UCT_fig:4}
\end{figure*}
\begin{multline}
\frac{{f_{\rm v}}^2}{4m^2}(E_{p'}-E_p)^2 \bar u(\vec p^{\, \prime})[\gamma_0 \gamma_{\nu}-g_{0 \nu}]u(\vec p) \\
\times  \bar u(-\vec p^{\, \prime})[\gamma^0 \gamma^{\nu}-g^{0 \nu}]u(-\vec p)
\label{Eq.70}
\end{multline}
in \eqref{Eq.69}, we obtain approximate expressions that with the common factor
\begin{equation*}
(2\pi )^{-3} m^2/E_{p'}E_{p}
\end{equation*}
instead of
\begin{equation*}
(2\pi)^{-3}m/\sqrt{E_{p'} E_{p}}
\end{equation*}
are equivalent to Eqs.(E.21)--(E.23) from \cite{MachHolElst87}.
Such an equivalence becomes coincidence if in our formulae instead of the canonical two-nucleon basis  $\left| {\vec p\,\mu _1 \mu _2 } \right\rangle $ one uses the helicity basis as in \cite{MachHolElst87}.

In parallel, we have considered the set of equations
\begin{figure*}[t]
\begin{center}
  \includegraphics[scale=0.80]{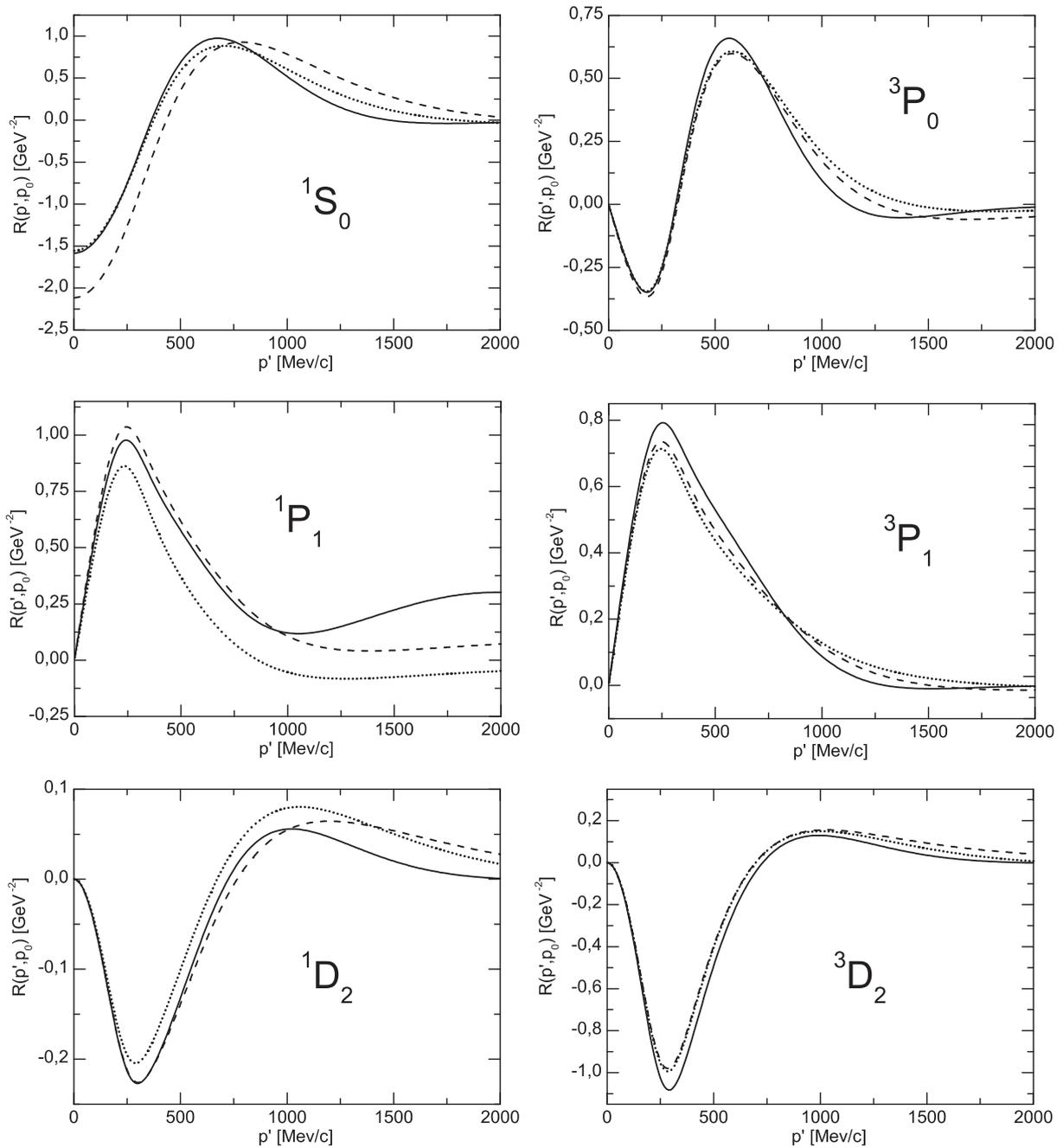}
\end{center}

\caption {Half--off--shell $R$--matrices for uncoupled waves at
laboratory energy equal to 150 MeV($p_0$=265 MeV). Other notations
as in Fig.3} \label{fig:5}
\end{figure*}
\begin{multline}
{^B}R_{l'{\rm{ }}l}^{JST} (p',p) = {^B} V_{l'{\rm{ }}l}^{JST}(p',p) \\
+ m \sum \limits_{l''} \int \limits_0^\infty \frac{d q}{p^2 - q^2} \left\{ q^2\, {^B}V_{l'l''}^{JST}(p',q) {^B}R_{l''l}^{JST}(q,p) \right. \\
\left. -  p^2\,{^B}V_{l'l''}^{JST}(p',p) {^B}R_{l''l}^{JST}(p,p) \right\} , \label{Eq.71}
\end{multline}
where the superscript $B$ refers to the partial matrix elements of the
potential $B$ determined in \cite{Mach89} with the just mentioned interchange of the bases. It is
important to note that
Eqs.\eqref{Eq.71} can be obtained from Eqs.\eqref{Eq.66} ignoring some relativistic effects.
In particular, it means that the covariant OBE propagators
\begin{equation}
\frac{1}{{(p' - p)^2  - m_b^2 }} = \frac{1}{(E_{p'}-E_p )^2 - (\vec p' -
\vec p)^2 - m_b^2 } \label{Eq.72}
\end{equation}
are replaced by their nonrelativistic counterparts
\begin{equation}
- \frac{1}{ (\vec p' - \vec p)^2 + m_b^2 } \label{Eq.73}
\end{equation}

Such an approximation \footnote{Sometimes associated with ignoring
the so-called meson retardation (see, e.g., Appendix E from
\cite{MachHolElst87}
and a discussion therein)} is a key point that gives rise to the potential
$B$ from \cite{Mach89}.

\begin{figure*}[!ht]
\begin{center}
  \includegraphics[scale=0.80]{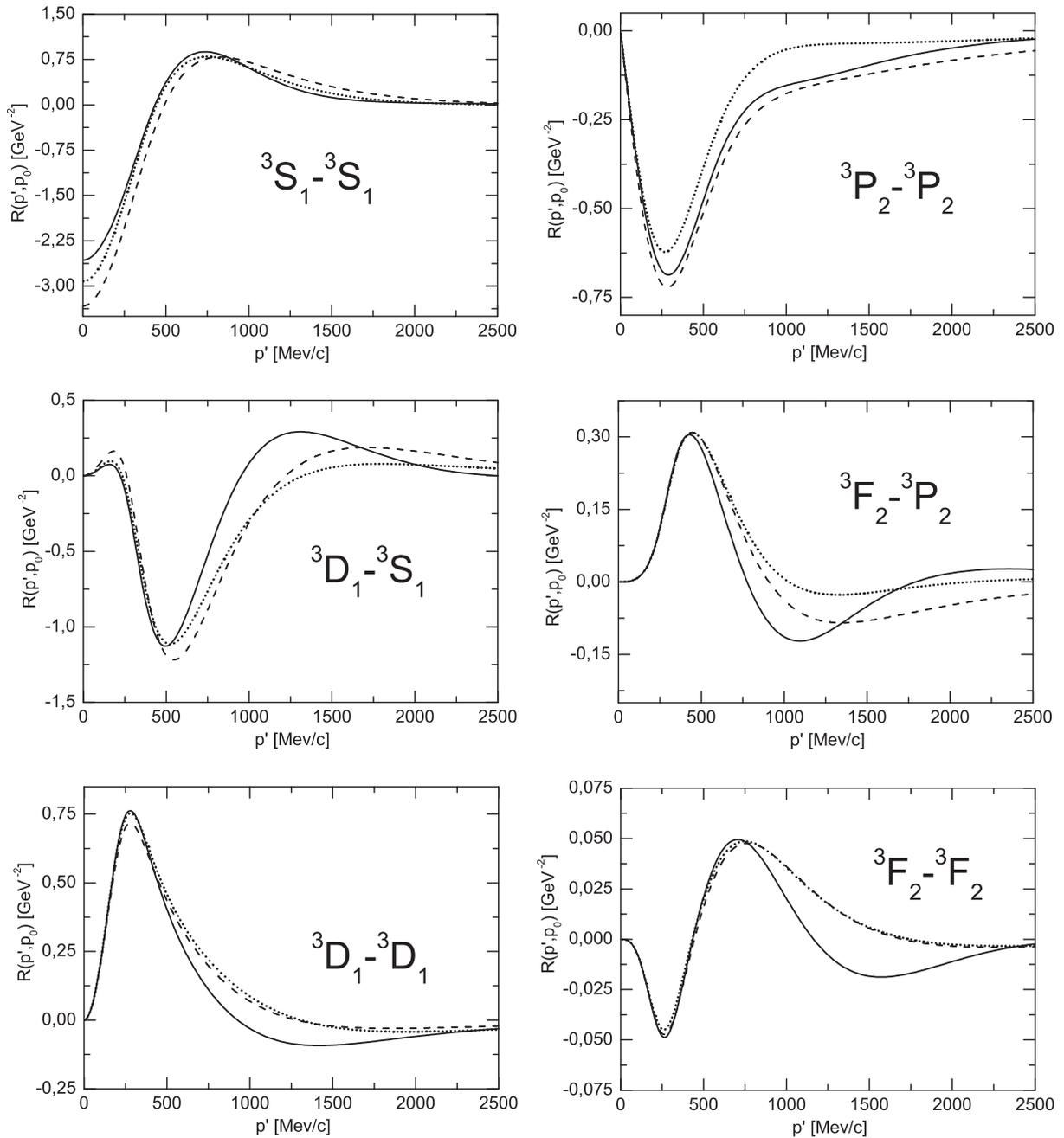}
\end{center}

\caption{The same in Fig.5 but for the coupled waves.}

\label{fig:6}
\end{figure*}

\begin{figure*}[t]
\begin{center}
  \includegraphics[scale=0.80]{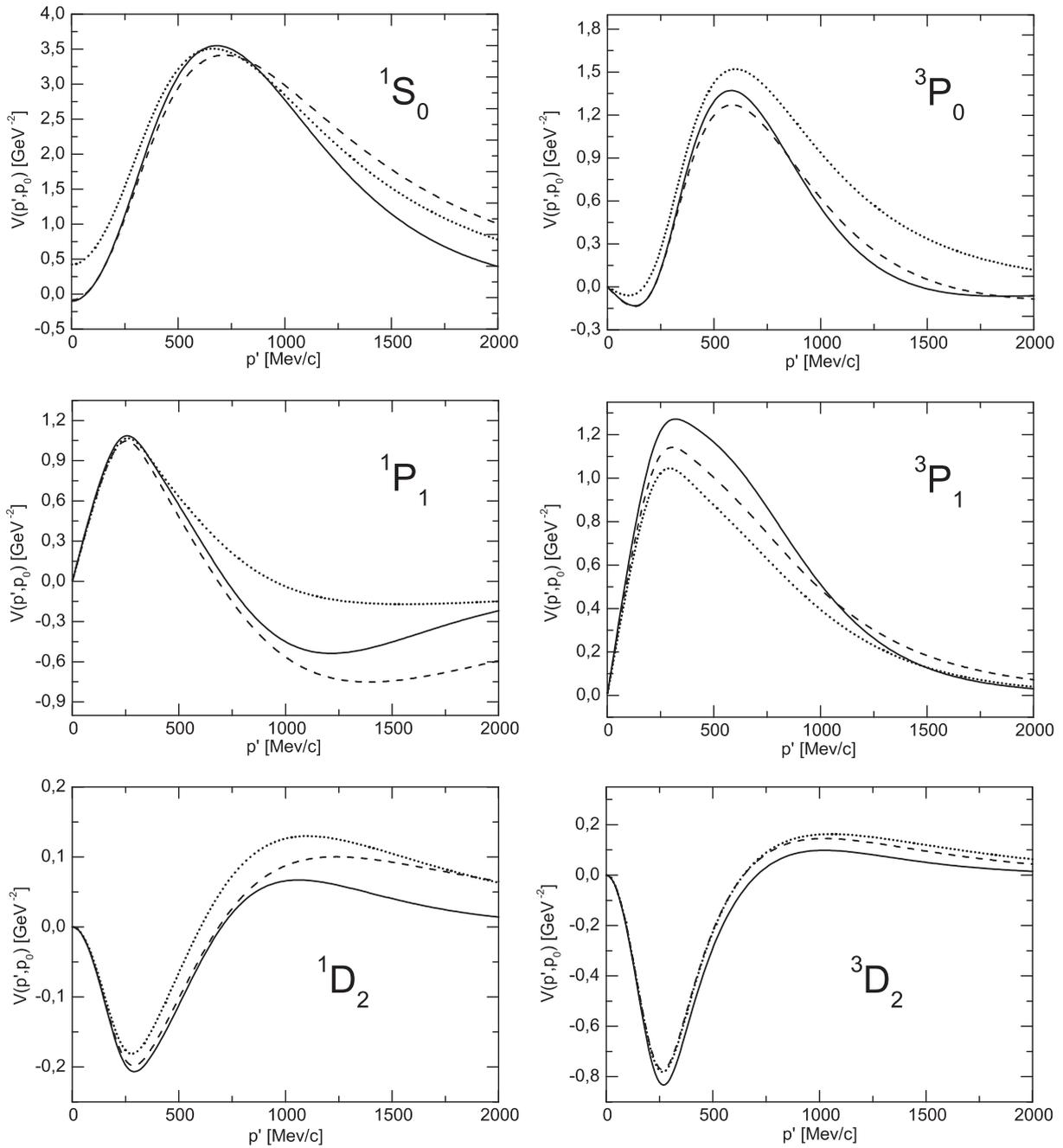}
\end{center}

\caption{Potentials for the uncoupled
         partial waves with the momentum $p_0$ fixed as in Fig. 5. Other notations in Fig. 3.}

\label{fig:7}
\end{figure*}

\begin{figure*}[t]
\begin{center}
  \includegraphics[scale=0.80]{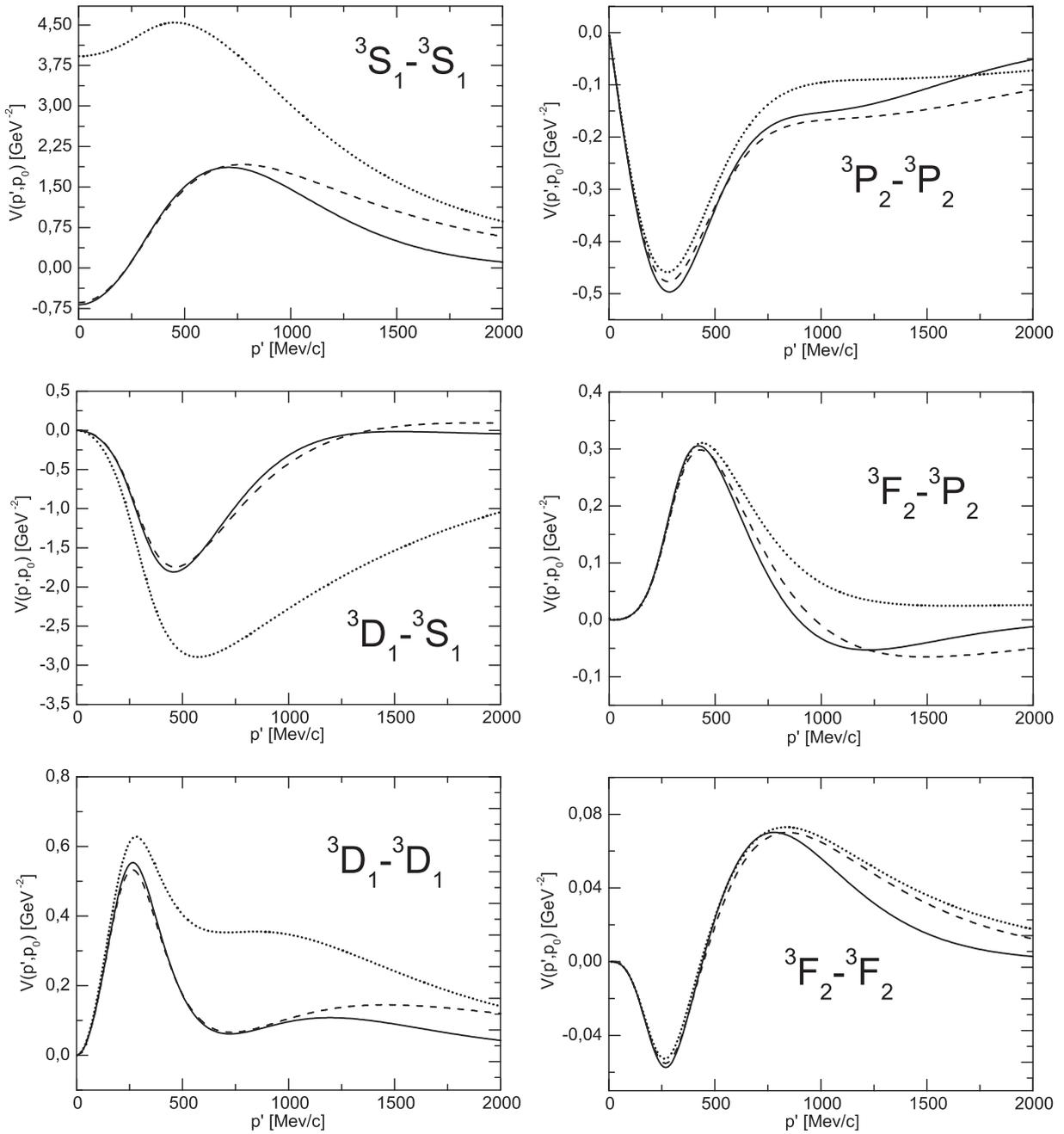}
\end{center}

\caption{The same as in Fig.7 for the coupled waves.}

\label{UCT_fig:8}
\end{figure*}

However, the transition from \eqref{Eq.72} to \eqref{Eq.73}, being valid on the energy shell
\begin{equation}
E_{p'} = E_p = \frac12 E_{cms},
\end{equation}
cannot be {\it a priori} justified when finding the $R$ matrix
even on the shell. Thus our calculations of the $R$ matrices that
meet the equations \eqref{Eq.66} and \eqref{Eq.71} are twofold. On
the one hand, we will check reliability of our numerical procedure
(in particular, its code). On the other hand, we would like to
show similarities and discrepancies between our results and those
by the Bonn group both on the energy shell and beyond it. These
results are depicted in Figs. \ref{UCT_fig:3}--\ref{UCT_fig:8} and
collected in Table \ref{UCT_tab:2}.

As seen in Figs. \ref{UCT_fig:3}--\ref{UCT_fig:4}, the most appreciable distinctions between
the UCT and OBEP curves take place for the phase shifts with the
lowest $l-$values. As the orbital angular momentum increases the
difference between the solid and da\-shed curves decreases. Such
features may be explained if one takes
into account that the approximations under consideration affect mainly high--momentum
components of the UCT quasipotentials (their behavior at "small"
distan\-ces). With the $l$--increase the influence of small
distances is suppressed by the centrifugal barrier repulsion.

Of course, it would be more instructive to compare the
corresponding half--off--energy--shell $R$--matrices (see
definition \eqref{Eq.51}). Their $p'$--dependencies shown in
Fig.5--6 are necessary to know when calculating the $\psi^{(\pm)}$
scattering states for a two--nucleon system.
Such states may be expressed through the partial--wave functions
$\varphi^a_{ll'} \,\, a={J,S,T}$ that have the asymptotic of
standing waves (see, for example, \cite{KorchinShebeko90}). Within
the MIM every $\varphi$ can be represented as
\begin{equation*}
\varphi^a_{ll'}(p) = \sum \limits_{j=1}^{N+1} B^a_{ll'}(j)\delta(p-p_j)/p^2_j,
\end{equation*}
where the coefficients $B^a_{ll'}(j)$ are the solutions of the set
of linear algebraic equations
approximately equivalent to the
integral equation for the corresponding $R$--matrix; $N$ is the
dimension of the set, $p_j$ are the grid points associated with
the Gaussian nodes on the interval $[-1,1]$, $p_{N+1}=p_0$
(details can be found in \cite{KorchinShebeko90,LadyShe04}).
Meanwhile, our computations presented here have been done with
$N=32$ (we do not talk about tests with other $N$-values to get
the results stable with respect to $N$).

Completing the comparison of the UCT and Bonn results we display
in Figs. 7--8 some interactions in relevant partial states.
Appreciable distinctions between dotted and solid curves in these
figures mean that the UCT and Bonn parameters from Table 1,
ensuring a fair treatment of such on--energy--shell quantities as
the phase shifts, may be inadequate in constructing model
nucleon--nucleon potentials. It seems to be especially prominent
in case of the $^3S_1-^3D_1$ potentials responsible for the
formation of the tensor part of nuclear forces.

In addition, one should emphasize that hitherto we have explored
the OBEP and UCT $R$--matrices in the c.m.s., where the both
approaches yield the most close results. It is not the case in
those situations when the c.m.s. cannot be referred to everywhere
(e.g., in the reactions $NN \rightarrow \gamma NN$ and $\gamma d
\rightarrow pn$). In this respect our studies of the differences
between UCT and OBE approaches are under way.

\section{Summary, Conclusions and Prospects}

The present work has been made to develop a consistent field-theoretical
approach in the theory of nucleon-nucleon scattering.
It has been shown that the method of UCT's, based upon the notion of
clothed particles, is proved to be appropriate in achieving this purpose.

Starting from a primary Lagrangian for interacting meson and
nucleon fields, we come to the corresponding Hamiltonian whose
interaction part $K_I$ consists of new relativistic interactions
responsible for physical (not virtual) processes in the system of
the bosons ($\pi-$, $\eta-$, $\rho-$, $\omega-$, $\delta-$
and $\sigma-$mesons) and the nucleon. Proceeding with the CPR we
have confined ourselves to constructing the four-legs interaction
operators $K_I^{(2)}$ in the two-nucleon sector of the Hilbert
space of hadronic states. The corresponding quasipotentials (these
essentially nonlocal objects) for binary processes $NN \rightarrow
NN$, $\bar NN \rightarrow \bar NN$, etc. are Hermitian and
energy--inde\-pendent. It makes them attractive for various
applications in nuclear phy\-sics. They embody the off--shell
effects in a natural way without addressing to any off--shell
extrapolations of the $S-$matrix for the $NN$ scattering.

Using the unitary equivalence of the CPR to the BPR, we have seen
how in the approximation $K_I = K_I^{(2)}$ the extremely
complicated scattering problem in QFT can be reduced to the three
--dimensional $LS$--type equation for the $T-$matrix in
momentum space. The equation kernel is given by the clothed
two-nucleon interaction of the class [2.2]. Such a conversion
becomes possible owing to the property of $ K_I^{(2)}$ to leave
the two--nucleon sector and its separate subsectors to be
invariant.

Special attention has been paid to the elimination of auxiliary field
components. We encounter such a necessity for interacting vector
and fermion fields when in accordance with the canonical formalism
the interaction Hamiltonian density embodies not  only a scalar
contribution but nonscalar terms too. It has proved (at least, for
the primary $\rho N$ and $\omega N$ couplings) that the UCT method
allows us to remove such noncovariant terms directly in the
Hamiltonian. To what extent this result will take place in higher orders in coupling constants
it will be a subject of further explorations.

Being concerned with constructing the two--nucleon
states from $\mathcal{H}$ and their angular--momentum
decomposition we have not used the so--called separable ansatz,
where every such state is a direct product of the corresponding
one-- nucleon (particle) states. The clothed two--nucleon partial
waves have been built up as common eigenstates of the field total
angular--momentum generator and its polarization (fermionic) part
expressed through the clothed creation/destruction operators and
their derivatives in momentum space.

We have not tried to attain a global treatment of modern precision data.
But a fair agreement with the earlier analysis by the Bonn group makes
sure that our approach may be useful for a more advanced analysis. In the
context, to have a more convincing argumentation one needs to do at least
the following. First, show the low-energy scattering parameters and the deuteron wave
function calculated within the UCT method. Second, consider triple
commutators
$[R,[R,[R,V_b]]]$ to extract the two--boson--two--nucleon interaction
operators of the same class [2.2] in the fourth order in the coupling
constants.
Third, extend our approach for describing the $NN$ scattering above the
pion production threshold. All the things are in progress.

As a whole, the persistent clouds of virtual particles are no
longer explicitly contained in the CPR, and their influence is
included in the properties of the clothed particles (these
quasiparticles of the UCT method). In addition, we would like to
stress that the problem of the mass and vertex renormalizations is
intimately interwoven with constructing the interactions between
the clothed nucleons. The renormalized quantities are calculated
step by step in the course of the clothing procedure unlike some
approaches, where they are introduced by "hands".

\section{Acknowledgements}

We are thankful to V.Korda, who has allowed us to use an original
code based upon a genetic algorithm for our best--fit procedure.

\appendix

\section{  Model Lagrangians and Hamiltonians within Canonical Formalism. Transition to
Clothed Particle Representation}

We will focused upon three boson fields (pseudoscalar, scalar and vector) that are coupled with
the nucleon by means of the often employed interaction Lagrangians,
\begin{equation}
\mathscr{L}_s = - g_s^0 \bar \Psi \Psi \Phi_s, \label{(A.1)}
\end{equation}
\begin{equation}
\mathscr{L}_{ps} = - ig_{ps}^0 \bar \Psi \gamma_5 \Psi \Phi_{ps}, \label{(A.2)}
\end{equation}
\begin{equation}
\mathscr{L}_{\rm{v}} = - g_{\rm{v}}^0 \bar \Psi \gamma_{\mu} \Psi \Phi_{\rm{v}}^{\mu}
-\frac{f_{\rm{v}}^0}{4m}\bar \Psi \sigma_{\mu \nu} \Psi \Phi_{\rm{v}}^{\mu \nu},  \label{(A.3)}
\end{equation}
where $\sigma_{\mu \nu}=\frac{i}{2}(\gamma_{\mu}\gamma_{\nu}-\gamma_{\nu}\gamma_{\mu})$ and $\Phi_{\rm{v}}^{\mu \nu}
=\partial^{\mu}\Phi_{\rm{v}}^{\nu}-\partial^{\nu}\Phi_{\rm{v}}^{\mu}$. As in Refs. (\cite{SheShi01}, \cite{KorCanShe07}),
throughout this paper we use the definitions and notations of \cite{BD}, so, e.g., $\gamma_{\mu}\gamma_{\nu}+\gamma_{\nu}\gamma_{\mu}=2g_{\mu \nu}$
($g_{00}=1$, $g_{11}=g_{22}=g_{33}=-1$), $\gamma_{\mu}^{\dag}=\gamma_0 \gamma_{\mu} \gamma_0$, $\gamma_5=i\gamma^0 \gamma^1 \gamma^2 \gamma^3$.
In addition, following Ref. \cite{WeinbergBook1995} we shall distinguish via upper(lower) case letters between the Heisenberg and Dirac picture field operators
($\Psi$ and $\Phi$ vs $\psi$ and $\varphi$, respectively, for fermions and bosons).

Of course, we could incorporate the so--called pseudovector (pv) coupling
\begin{equation}
\mathscr{L}_{pv} = - \frac{f_{pv}^0}{m_{ps}} \bar \Psi \gamma^5 \gamma^{\mu} \Psi \partial_{\mu} \Phi_{ps}.  \label{(A.4)}
\end{equation}
Since in this paper all used model interactions are suggested to be modified by introducing some cutoff factors, it has no matter that couplings \eqref{(A.3)} and \eqref{(A.4)}
with derivatives are nonrenormalizable (cf. an instructive discussion of this subject in Subsect 3.4 of the survey \cite{Mach89}). In the context, starting from couplings \eqref{(A.1)}--\eqref{(A.3)} with "bare" constants $g_{ps}^0$, $g_s^0$, $g_{\rm{v}}^0$ and $f_{\rm{v}}^0$, we have tried to reproduce some results obtained in Refs. \cite{MachHolElst87}, \cite{Mach01}, where such constants from the beginning are replaced by effective parameters $g_{ps}$, $g_s$, $g_{\rm{v}}$ and $f_{\rm{v}}$. It explains (at least, {\it ad hoc}) our restriction to these Lagrangian densities. Recall also that for isospin 1 bosons one needs to write $\vec \tau \vec \Phi_b$ instead of $\Phi_b$, where $\vec \tau$ is the Pauli vector in isospin space.

In constructing the Hamiltonians with Lagrangian densities \eqref{(A.1)}--\eqref{(A.3)} as a departure point, we have first used the equations of motion for the $\rm{H}$ fields
and the so--called Legendre transformation (from $L$ to $H$) to express the total Hamiltonian $H$ in terms of the independent canonical variables and their conjugates. Then, passing to the D picture (interaction representation) the Hamiltonian has been split into a physically satisfactory free--field part $H_0$ and an interaction $V$. {\it Int. al.}, since the component $\Phi_{\rm{v}}^0$ has no canonical conjugate, we have resorted to a trick prompted by Eq.(7.5.22) from \cite{WeinbergBook1995} to introduce a proper component $\varphi_{\rm{v}}^0$ in the Dirac picture. As a result, we arrive to the interaction Hamiltonian densities:
\begin{equation}
\mathscr{H}_s(x) = g_s^0 \bar \psi(x) \psi(x) \varphi_s(x), \label{(A.5)}
\end{equation}
\begin{equation}
\mathscr{H}_{ps}(x) = ig_{ps}^0 \bar \psi(x) \gamma_5 \psi(x) \varphi_{ps}(x), 
\end{equation}
\begin{equation}
\mathscr{H}_{\rm{v}}(x) = \mathscr{H}_{\rm{cov}}(x) + \mathscr{H}_{\rm{ncov}}(x), \label{(A.7)}
\end{equation}
where
\begin{equation}
\mathscr{H}_{\rm{cov}}(x) = g_{\rm{v}}^0 \bar \psi(x) \gamma_{\mu} \psi(x) \varphi_{\rm{v}}^{\mu}(x)
+\frac{f_{\rm{v}}^0}{4m}\bar \psi(x) \sigma_{\mu \nu} \psi(x) \varphi_{\rm{v}}^{\mu \nu}(x)  
\end{equation}
and
\begin{multline}
\mathscr{H}_{\rm{ncov}}(x) = \frac{g_{\rm{v}}^{0\,2}}{2m_{\rm{v}}^2}\bar \psi(x) \gamma_0 \psi(x) \bar \psi(x) \gamma_0 \psi(x) \\
+\frac{f_{\rm{v}}^{0\,2}}{4m^2}\bar \psi(x) \sigma_{0i} \psi(x)\bar \psi(x) \sigma_{0i} \psi(x).  
\end{multline}
It is implied that the total Hamiltonian of interest $H=H_0+V$ consists of the sum $H_0=H_{0,f}+\sum \limits_b H_{0,b}$, where $H_{0,f}$ the free--fermion Hamiltonian,
$H_{0,b}$ separate free--boson contribution, and the space integral
\begin{equation}
V = \int d \vec x \mathscr{H}(\vec x)   \label{(A.10)}
\end{equation}
of the interaction density $\mathscr{H}(x)$ in the D picture
\begin{equation}
\mathscr{H}(x) = \mathscr{H}_{s}(x) + \mathscr{H}_{ps}(x) + \mathscr{H}_{\rm{v}}(x),  
\end{equation}
taken at $t=0$, i.e., in the Schr\"{o}dinger (S) picture. In other words, $\mathscr{H}(\vec x)=\mathscr{H}(0,\vec x)$.

Expressions \eqref{(A.5)}-\eqref{(A.10)} exemplify that for a Lorentz--invariant Lagrangian it is not necessarily to have "... the interaction Hamiltonian as the integral over space of a scalar interaction density; we also need to add non--scalar terms to the interaction density ..." (quoted from p.292 of Ref. \cite{WeinbergBook1995}). It is the case with derivative couplings and/or spin $\geq 1$.

In this connection, let us recall the property of the density $\mathscr{H}(x)$ to be scalar, viz.,
\begin{equation}
U_F(\Lambda, a) \mathscr{H}(x) U_F^{-1}(\Lambda, a) = \mathscr{H}(\Lambda x + a)  ,  
\end{equation}
where the operators $U_F(\Lambda, a)$ realize a unitary irreducible representation of the Poincar\'{e} group in the Hilbert space of states
for free (non--interacting) fields. Other comments will be given in Appendix B.

Moreover, it is well known that this property is a key point of covariant perturbation theory of the $S$--matrix (see, e.g., Chapter V of \cite{WeinbergBook1995} and refs. therein). In the respect, the division \eqref{(A.7)} is not accidental.

Further, the BPR form \eqref{Eq.1} and its comparison with Eq. \eqref{(A.10)} give rise to the interactions \eqref{Eq.3}--\eqref{Eq.5} between the bare bosons and fermions with physical masses.
The next step is to apply the clothing procedure exposed in \cite{SheShi01} (see also \cite{KorCanShe07}), where the first clothing transformation $W^{(1)}=\exp[R^{(1)}]$ (${R^{(1)}}^{\dag}=-R^{(1)}$) eliminates all interactions linear in the coupling constants. Its generator $R^{(1)}$ obeys the equation
\begin{equation}
\left[ R^{(1)}, H_F \right] + V^{(1)} = 0    \label{(A.13)}
\end{equation}
with
\begin{equation*}
V^{(1)} = V_s + V_{ps} + V_{\rm{v}}^{(1)},
\end{equation*}
where
\begin{multline}
V_{\rm{v}}^{(1)} = \int d \vec x \left\{ g_{\rm{v}} \bar \psi(\vec x) \gamma_{\mu} \psi(\vec x) \varphi_{\rm{v}}^{\mu}(\vec x)
\phantom{\frac{f_{\rm{v}}}{4m}} \right. \\
\left. +\frac{f_{\rm{v}}}{4m}\bar \psi(\vec x) \sigma_{\mu \nu} \psi(\vec x) \varphi_{\rm{v}}^{\mu \nu}(\vec x) \right\}.  
\end{multline}
Following \cite{SheShi01} Eq.\eqref{(A.13)} is satisfied with
\begin{equation}
R^{(1)} = -i\lim \limits_{\varepsilon \rightarrow 0+} \int \limits_0^{\infty} V_D^{(1)}(t)e^{-\varepsilon t} dt    
\end{equation}
if $m_b < 2m$.

The corresponding interaction operator in the CPR (see relation \eqref{K(2)general_structure} of the text) can be written as
\begin{equation}
K_I^{(2)} = \frac{1}{2}\left[ R^{(1)}, V^{(1)} \right] + V^{(2)},   \label{(A.16)}
\end{equation}
where we have kept only the contributions of the second order in coupling constants, so
\begin{multline}
V^{(2)} = \int d \vec x \left\{  \frac{g_{\rm{v}}^{2}}{2m_{\rm{v}}^2}\bar \psi(x\vec ) \gamma_0 \psi(\vec x) \bar \psi(\vec x) \gamma_0 \psi(\vec x) \right. \\
\left. +\frac{f_{\rm{v}}^{2}}{4m^2}\bar \psi(\vec x) \sigma_{0i} \psi(\vec x)\bar \psi(\vec x) \sigma_{0i} \psi(\vec x)  \right\}     \label{(A.17)}
\end{multline}
that coincides with the space integral of the non--scalar density $\mathscr{H}_{ncov}(x)$ at $t=0$.
It is implied that all operators in the r.h.s. of Eq.\eqref{(A.16)} depend on the clothed--particle creation(destruction) operators $\alpha_c = {W^{(1)}}^{\dag} \alpha W^{(1)}$,
for example, involved in the Fourier expansions
\begin{multline*}
\psi(\vec x) = (2\pi)^{-3/2} \int d \vec p \sqrt{\frac{m}{E_{\vec p}}} \sum \limits_{\mu} \left[u (\vec p \mu) b_c(\vec p \mu) \right. \\
\left. + v(-\vec p \mu) d_c^{\dag}(-\vec p \mu) \right] \exp (i \vec p \vec x)  ,
\end{multline*}
\begin{multline*}
\varphi_{\rm{v}}^{\mu}(\vec x) = (2\pi)^{-3/2} \int  \frac{d \vec k}{\sqrt{2\omega_{\vec k}}} \sum \limits_{s}  \left[ e^{\mu}(\vec k,s) a_c(\vec k,s) \right. \\
\left. + e^{\mu}(- \vec k,s) a_c^{\dag}(-\vec k,s) \right] \exp (i \vec k \vec x),
\end{multline*}
where the $e^{\mu}(\vec k,s)$ for $s=+1,0,-1$ are three independent vectors, being transverse $k_{\mu}e^{\mu}(\vec k,s)=0$, and normalized so that
\begin{equation*}
\sum \limits_{s} e_{\mu}(\vec k,s) e_{\nu}^{*}(\vec k,s) = -g_{\mu \nu} + k_{\mu}k_{\nu}/m^2  .
\end{equation*}
In this paper we do not intend to derive all interactions between the clothed mesons and nucleons, allowed by formula \eqref{(A.16)}. Our aim is more humble, viz., to find in the r.h.s. of Eq.\eqref{(A.16)} terms of the type \eqref{Eq.18}, responsible for the N--N interaction. Along the guideline the commutators $\frac{1}{2}[R^{(1)}_{s},V_{s}]$ and $\frac{1}{2}[R^{(1)}_{ps},V_{ps}]$ generate the scalar-- and pseudoscalar--meson contributions $K_b(NN \rightarrow NN)$ with coefficients \eqref{Eq.20}--\eqref{Eq.21}. In case of the vector mesons we encounter an interplay between the commutator $\frac{1}{2}[R^{(1)}_{\rm{v}},V_{\rm{v}}^{(1)}]$ and the integral \eqref{(A.17)}.

To show it explicitly, let us write
\begin{multline}
V_{\rm{v}}^{(1)} =  - \frac{1}{( 2\pi)^{3/2}} \sum \limits_{\mu s} \int d \vec k d \vec p' d \vec p
\frac{m}{\sqrt {2\omega_{\vec k} E_{\vec p'} E_{\vec p}}} e^\rho (\vec k,s) \\
\times \delta \left( \vec p^{\,\prime} - \vec p - \vec k \right) \bar u( \vec p^{\,\prime} \mu')\left\{ {g_{\rm{v}}\gamma _\rho   - \frac{f_{\rm{v}}}{2m}i\sigma _{\nu \rho } k^\nu  } \right\} u( \vec p \mu) \\
\times b_c^\dag ( \vec p' \mu') b_c(\vec p \mu) a_c( \vec k,s) + {\rm{H}}{\rm{.c}}.,   
\end{multline}
\begin{multline}
R_{\rm{v}}^{(1)}  =  - \frac{1}{(2\pi)^{3/2}} \sum  \limits_{\mu s} \int d \vec k d \vec p' d \vec p
\frac{m}{\sqrt{2\omega_{\vec k} E_{\vec p'} E_{\vec p}}} e^\rho  ( \vec k,s)  \\
\times \frac{\delta \left( {\vec p^{\,\prime} - \vec p - \vec k} \right)}{E_{\vec p'}  - E_{\vec p}  - \omega_{\vec k}}
\bar u( \vec p' \mu') \left\{ g_{\rm{v}}\gamma _\rho   - \frac{f_{\rm{v}}}{2m}i\sigma _{\nu \rho } k^\nu   \right\} u( \vec p \mu) \\
\times b_c^\dag ( \vec p' \mu') b_c( \vec p \mu) a_c( \vec k,s) - {\rm{H}}{\rm{.c}}.,     
\end{multline}
retaining only those parts of $V^{(1)}$ and $R^{(1)}$, which are necessary for deriving $K_{\rm{v}}(NN \rightarrow NN)$. After a simple algebra we find
\begin{multline*}
\frac12 \left[ R^{(1)}, V^{(1)} \right]_{\rm{v}}(NN \rightarrow NN) \\
=K_{\rm{v}}(NN \rightarrow NN) + K_{cont}(NN \rightarrow NN)
\end{multline*}
with
\begin{multline*}
K_{\rm{v}}(NN \rightarrow NN) =
\frac{1}{2(2\pi)^3} \sum \limits_{\mu} \int d \vec p'_1 d \vec p'_2 d \vec p_1 d \vec p_2 \\
\times \frac{m^2}{\sqrt {E_{\vec p'_1 } E_{\vec p'_2 } E_{\vec p_1 } E_{\vec p_2 }}}
\frac{ \delta \left( {\vec p^{\,\prime}_1  + \vec p^{\,\prime}_2  - \vec p_1  - \vec p_2 } \right)}{(p_1^{\,\prime}  - p_1)^2  - m_b^2 } \\
\times \left[ \bar u(\vec p_1^{\,\prime} \mu'_1) \left\{ (g_{\rm{v}} + f_{\rm{v}})\gamma_\nu - \frac{f_{\rm{v}}}{2m}(p'_1  + p_1)_\nu  \right\} u(\vec p_1\mu_1) \right. \\
\shoveright{ \times \bar u(\vec p_2^{\,\prime} \mu'_2)\left\{ (g_{\rm{v}} + f_{\rm{v}})\gamma ^\nu - \frac{f_{\rm{v}}}{2m}(p'_2 + p_2)^\nu \right\} u(\vec p_2 \mu_2) } \\
- (E_{p'_1 } + E_{p'_2 }  - E_{p_1 }  - E_{p_2 }) \\
\times \bar u(\vec p_1^{\,\prime} \mu'_1) \left\{ (g_{\rm{v}} + f_{\rm{v}}) \gamma _\nu - \frac{f_{\rm{v}}}{2m}(p'_1 + p_1)_\nu  \right\} u(\vec p_1 \mu_1) \\
\shoveright{ \left. \times \bar u(\vec p_2^{\,\prime} \mu'_2) \frac{f_{\rm{v}}}{2m} \left\{ \gamma ^0 \gamma ^\nu   - g^{0\nu} \right\} u(\vec p_2 \mu_2) \right] } \\
\times b_c^\dag (\vec p^{\,\prime}_1 \mu'_1) b_c^\dag (\vec p'_2 \mu'_2) b_c (\vec p_1 \mu_1) b_c (\vec p_2 \mu_2)
\end{multline*}
and
\begin{multline*}
K_{cont}(NN \rightarrow NN) =
\frac{1}{2(2\pi)^3} \sum \limits_{\mu} \int d \vec p'_1 d \vec p'_2 d \vec p_1 d \vec p_2 \\
\times \frac{m^2}{\sqrt {E_{\vec p'_1 } E_{\vec p'_2 } E_{\vec p_1 } E_{\vec p_2 }}}
\delta \left( \vec p^{\,\prime}_1  + \vec p^{\,\prime}_2  - \vec p_1  - \vec p_2 \right) \\
\times \left[ \frac{g_{\rm{v}}^2}{m_{\rm{v}}^2} \bar u(\vec p^{\,\prime}_1 \mu'_1) \gamma _0
u(\vec p_1 \mu_1) \bar u(\vec p^{\,\prime}_2 \mu'_2)\gamma _0 u(\vec p_2 \mu_2) \right. \\
\shoveright{ \left. - \frac{f_{\rm{v}}^2}{4m^2}\bar u(\vec p^{\,\prime}_1 \mu'_1) \gamma _0 \vec \gamma u(\vec p_1 \mu_1) \bar u(\vec p^{\,\prime}_2 \mu'_2) \gamma _0 \vec \gamma  u(\vec p_2 \mu_2)  \right] } \\
\times b_c^\dag (\vec p^{\,\prime}_1 \mu'_1) b_c^\dag (\vec p'_2 \mu'_2) b_c (\vec p_1 \mu_1) b_c (\vec p_2 \mu_2)
\end{multline*}
The latter may be associated with a contact interaction since it
does  not contain any propagators (cf. the approach by the Osaka
group \cite{TamuraSato88}). It is easily seen that this operator
cancels completely the non--scalar operator $V^{(2)}$. In other
words the first UCT enables us to remove the non--invariant terms
directly in the Hamiltonian. It gives an opportunity to work with
the Lorentz scalar interaction only (at least, in the second order
in the coupling constants). In our opinion, such a cancellation,
first discussed here, is a pleasant feature of the CPR.

The remaining vector--meson contribution $K_{\rm{v}}(NN
\rightarrow NN)$ is  determined by coefficients \eqref{Eq.22} and
gives us one more relativistic interaction in the boson--fermion
system under consideration.

\section{  Partial Wave Expansion of Two--Particles States in the CPR}

We will show how one can proceed without the separable ansatz
\begin{equation*}
\left|{\vec p_1 \vec p_2 \mu _1 \mu _2 } \right\rangle  = \left| {\vec
p_1 \mu _1 } \right\rangle \left| {\vec p_2 \mu _2 } \right\rangle
\end{equation*}
often exploited in relativistic quantum mechanics (RQM) (see,
e.g., \cite{Werle66} and \cite{KeisPoly91}) in getting expansions
similar to Eq.(47). Unlike this, our consideration with particle
creation/destruction as a milestone, where the clothed
two--nucleon state is given by~ \footnote{Sometimes it is
convenient to handle the operators
$b_c^{\dag}(p\mu)=\sqrt{p_0}b_c^{\dag}(\vec{p}\mu)$ and their
adjoint ones $b_c(p\mu)$ that meet covariant relations $\left\{
{b_c^\dag (p'\mu '),b_c(p\mu )} \right\} = p_0 \delta (\vec p' -
\vec p)\delta_{\mu' \mu}$ }
\begin{equation}
\left| {\vec p_1 \mu _1 ;\vec p_2 \mu _2 } \right\rangle  =
b_c^\dag  (p_1 \mu _1 )b_c^\dag  (p_2 \mu _2 )\left| \Omega
\right\rangle
\end{equation}
By definition, it belongs to the two--nucleon sector of
$\mathscr{H}$ being the $K_F$ eigenstate with energy
$E=p_1^0+p_2^0$. Moreover, it is assumed that vector $ \left|
\Omega  \right\rangle $, being the single clothed no--particle
state, has the property
\begin{equation}
U_F (\Lambda ,a)\left| \Omega  \right\rangle  = \mathscr{P}\left|
\Omega  \right\rangle  =  \ldots  = \left| \Omega  \right\rangle
\end{equation}
\begin{equation*}
\forall \,\, \Lambda \in L_{+} \mbox{ and arbitrary spacetime shifts } a=(a^0,\vec a)
\end{equation*}
to be invariant with regard to the Poincar\'{e} group $\Pi$
\footnote{The correspondence $(\Lambda,a)\rightarrow
U_F(\Lambda,a)$ between elements $(\Lambda,a)\in \Pi$ and unitary
transformations $U_F(\Lambda,a)$ realizes an irreducible
representation of $\Pi$ on $\mathscr{H}$ (see, e.g., Chapter 2 in
\cite{WeinbergBook1995})}, space inversion and other symmetries.
Here $L_{+}$ is the homogeneous (proper) orthochronous Lorentz
group. In turn, every $U_F(\Lambda,a)$ is expressed through the
clothed free--particle generators of space--time translations
\begin{equation*}
P_F^\mu  (\alpha _c ) = \left\{ {K_F (\alpha _c ),\vec P_F (\alpha
_c )} \right\},
\end{equation*}
space rotations
\begin{equation*}
\vec J_F (\alpha _c ) = \left\{ {M_F^{23} (\alpha _c ),M_F^{31}
(\alpha _c ),M_F^{12} (\alpha _c )} \right\}
\end{equation*}
and the Lorentz boosts
\begin{equation*}
\vec B_F (\alpha _c ) = \left\{ {M_F^{01} (\alpha _c ),M_F^{02}
(\alpha _c ),M_F^{03} (\alpha _c )} \right\}
\end{equation*}
viz.,
\begin{equation}
U_F \left( {\Lambda ,a} \right) = \exp \left[ {ia_\rho  P_F^\rho
(\alpha _c )   + \frac{i}{2}\omega _{\rho \nu } M_F^{\rho \nu }
(\alpha _c ) } \right],    
\end{equation}
where the antisymmetric tensor $\omega_{\mu \nu}=-\omega_{\nu
\mu}$ with $(\mu,\nu=0,1,2,3)$ has six independent components.

Now, let us remind of the following transformation property:
\begin{equation}
U_F \left( {\Lambda ,a} \right)b_c^\dag  \left( {p\mu }
\right)U_F^{ - 1} \left( {\Lambda ,a} \right) = e^{i\Lambda p
\cdot a} b_c^\dag \left( {\Lambda p\mu '} \right) D_{\mu '\mu
}^{(1/2)} \left( {W\left( {\Lambda ,p} \right)} \right)  \label{(B.4)}
\end{equation}
with the $D^{(1/2)}$ function whose argument is the Wigner rotation
$W(\Lambda,p)$. The latter has the property $W(R,p)=R$ for any
three--dimensional rotation $R$. In other words, from Eq.\eqref{(B.4)} it
follows that under such a rotation, when
$U_F(\Lambda,a)=U_F(R,0)\equiv U_F(R)$, one has
\begin{equation}
U_F \left( {R} \right)b_c^\dag  \left( {p\mu } \right)U_F^{ - 1}
\left( {R} \right) =  b_c^\dag  \left( {R p \, \mu '} \right) D_{\mu
'\mu }^{(1/2)} \left( R \right)  
\end{equation}
In addition, we need to have an analytic expression for the
operator $\vec{J}_{ferm}$ to be expressed in terms of $b^{\dag}_c$
and $b_c$. To do it we recur to
the well known result:
\begin{equation}
\vec J_{ferm}  = \int {d \vec x\psi ^\dag  } (\vec x)\left[ {-i\vec x
\times \frac{\partial }{{\partial \vec x}} + \frac{1}{2}\vec
\Sigma } \right]\psi (\vec x)  \label{(B.6)}
\end{equation}
with $\vec \Sigma  = \left( {\begin{array}{*{20}c}
   {\vec \sigma } & 0  \\
   0 & {\vec \sigma }  \\
\end{array}} \right)$, where $\vec{\sigma}$ the Pauli vector.

After modest effort one can see that
\begin{equation}
\vec J_{ferm}  = \vec{L}_{ferm}+\vec{S}_{ferm},  
\end{equation}
where  $\vec{L}_{ferm}$ ($\vec{S}_{ferm}$) the orbital (spin)
momentum of the fermion field, that are given by
\begin{multline}
\vec{L}_{ferm}=\frac{i}{2}\sum\limits_\mu  {\int { d \vec p \,\,\vec p \times \left[ \frac {\partial b_c^\dag (\vec
p\mu )} {\partial \vec p} b_c(\vec p\mu ) - b_c^\dag  (\vec p\mu
)\frac{{\partial b_c(\vec p\mu )}}{{\partial \vec p}} \right. }} \\
\left. + \frac{\partial d_c^\dag  (\vec p\mu )}{\partial \vec
p}d_c(\vec p\mu ) - d_c^\dag  (\vec p\mu ) \frac{\partial d_c(\vec
p\mu )}{\partial \vec p} \right]   
\end{multline}
and
\begin{multline*}
\vec{S}_{ferm} = \frac12 \sum \limits_{\mu \mu'} \int d \vec p \,\frac{m}{E_{\vec p}} \left[ \left\{ u^\dag (\vec p, \mu') \vec \Sigma u(\vec p, \mu) \phantom{\frac12} \right.  \right. \\
- \left. i \vec p \times \left( u^\dag (\vec p, \mu') \frac{\partial u(\vec p, \mu)}{\partial \vec p} - \frac{\partial u^\dag  (\vec p, \mu')}{\partial \vec p}
u(\vec p, \mu) \right) \right\} b_c^\dag (\vec p \mu') b_c(\vec p \mu ) \\
- \left\{ \upsilon ^\dag  (\vec p, \mu') \vec \Sigma \upsilon (\vec p, \mu)
- i \vec p \times \left( \upsilon ^\dag  (\vec p, \mu') \frac{\partial \upsilon (\vec p, \mu)}{\partial \vec p} \phantom{\frac12} \right. \right. \\
\left. \left. \left.  - \frac{\partial \upsilon ^\dag (\vec p, \mu')}{\partial \vec p} \upsilon (\vec p, \mu) \right) \right\} d_c^\dag (\vec p \mu') d_c(\vec p \mu) \right],
\end{multline*}
or
\begin{equation}
\vec{S}_{ferm} = \frac{1}{2} \sum \limits_{\mu \mu'} \int  d \vec p \,\,  \chi^{\dag}_{\mu'} \vec \sigma \chi_{\mu}  \left \{ b_c^\dag (\vec p \mu ')b_c(\vec p \mu )
- d_c^\dag (\vec p \mu ')d_c(\vec p \mu ) \right \},    
\end{equation}
since
\begin{multline*}
{u^\dag  (\vec p,\mu ')\frac{{ \partial u(\vec p,\mu )}}{{\partial \vec p}} - \frac{{ \partial u^\dag  (\vec p,\mu ')}}{{\partial \vec p}}u(\vec p,\mu )} \\
 = {\upsilon ^\dag  (\vec p,\mu ')\frac{{ \partial \upsilon (\vec p,\mu )}}{{\partial \vec p}} - \frac{{ \partial \upsilon ^\dag  (\vec p,\mu ')}}{{\partial \vec p}}\upsilon (\vec p,\mu )} \\
 = \frac{i}{m(E_p+m)} \chi^{\dag}_{\mu'} \vec \sigma \chi_{\mu} \times \vec p,
\end{multline*}
if one uses the Dirac spinors defined in \cite{BD}. In these formulae we see the Pauli spinors $\chi_{\mu}$.
It is important to keep in mind that
\begin{equation*}
\vec{L}_{ferm}\neq -i \int {d^3 x\psi ^\dag  } (\vec x)  \vec x
\times \frac{\partial }{{\partial \vec x}} \psi (\vec x)
\end{equation*}
and
\begin{equation*}
\vec{S}_{ferm} \neq  \frac{1}{2} \int {d^3 x \psi ^\dag (\vec x) }
\vec \Sigma  \psi (\vec x)
\end{equation*}
In fact, the operator $\vec{S}_{ferm}$ stems from a destructive
interplay between orbital and spin parts of decomposition \eqref{(B.6)}.
In our opinion, such an interpretation differs from the definition
below Eq.(13.48) in \cite{BD}.

Hitherto, we have preserved the $\alpha_c$--dependence of the
relevant operators although, as shown in Sect. 3 of
\cite{SheShi01}, in the instant form of relativistic dynamics the
operator $\vec{P}_F(\alpha_c)(\vec{J}_F(\alpha_c))$ coincides with
the total linear momentum $\vec{P}\equiv \vec{P}_F(\alpha)$ (the
total angular momentum $\vec{J}\equiv \vec{J}_F(\alpha)$). Such an
observation allows us to omit the label $c$ if it does not lead to
confusion.

After these preliminaries we prefer to proceed sufficiently
straightforwardly repeating the well known steps (cf.
\cite{Werle66} and refs. therein). First, when handling the c.m.s.
two--nucleon state \footnote{Of course, what follows can be
extended to an arbitrary frame.}
\begin{equation}
\left| {\vec p\mu _1 \mu _2 } \right\rangle  = \left| {\vec p\mu
_1 ; - \vec p\mu _2 } \right\rangle   \label{(B.10)}
\end{equation}
let us consider the vector \footnote{For a moment, the isospin
quantum numbers are suppressed. We will come back to the point
later.}
\begin{equation}
\left| {\vec pSM_S } \right\rangle  =  {\left( { \left.
\frac{1}{2} \mu _1 \frac{1}{2} \mu _2 \right| SM_S } \right)}
\left| {\vec p\mu _1 \mu _2 } \right\rangle,   \label{(B.11)}
\end{equation}
so
\begin{equation}
\left| {\vec p\mu _1 \mu _2 } \right\rangle = {\left( { \left.
\frac{1}{2} \mu _1 \frac{1}{2} \mu _2 \right| SM_S } \right)}
\left| {\vec pSM_S } \right\rangle    \label{(B.12)}
\end{equation}
Second, one introduces
\begin{equation}
\left| {pJ(lS)M_J } \right\rangle = \int { d \hat{\vec{p}}  } \,
Y_{lm_l } \left( \hat{\vec{p}} \right) \left| {\vec p SM_S }
\right\rangle \left( {lm_l SM_S \left| {JM_J } \right.} \right)  
\end{equation}
or, reversely,
\begin{equation}
\left| {\vec pSM_S } \right\rangle  = \left( {lm_l SM_S \left|
{JM_J } \right.} \right) Y_{lm_l }^{*} \left( \hat{\vec{p}}
\right) \left| {pJ(lS)M_J } \right\rangle      \label{(B.14)}
\end{equation}
with the unit vector $\hat{\vec{p}} = \vec{p}/p$.

Third, substituting \eqref{(B.14)} into the r.h.s. of Eq.\eqref{(B.12)} we arrive
to the desired expansion,
\begin{multline}
\left| {\vec p\mu _1 \mu _2 } \right\rangle \\
= {\left( { \left. \frac{1}{2} \mu _1 \frac{1}{2} \mu _2 \right| SM_S } \right)}
\left( {lm_l SM_S \left| {JM_J } \right.} \right) Y_{lm_l }^{*}
\left( \hat{\vec{p}} \right) \left| {pJ(lS)M_J } \right\rangle    
\end{multline}
From the physical viewpoint it is important to know that
\begin{equation}
{\vec{J}}^{\,2} \left| {pJ(lS)M_J } \right\rangle = J(J+1) \left|
{pJ(lS)M_J } \right\rangle    
\end{equation}
\begin{equation*}
J^3 \left| {pJ(lS)M_J } \right\rangle = M_J \left| {pJ(lS)M_J }
\right\rangle
\end{equation*}
and
\begin{equation}
{\vec{S}}^{\,2}_{ferm} \left| {\vec pSM_S } \right\rangle = S(S+1)
\left| {\vec pSM_S } \right\rangle   \label{(B.17)}
\end{equation}
\begin{equation*}
S_{ferm}^3 \left| {\vec pSM_S } \right\rangle = M_S \left| {\vec
pSM_S } \right\rangle
\end{equation*}
A simple way of deriving these relations is to use the transformation
\begin{equation}
U_F(R) | \vec p S M_S \rangle = |R \vec p SM_S^{\prime} \rangle D_{M_S^{\prime}M_S}^{(S)}(R)    
\end{equation}
\begin{equation*}
\forall \, R \, \in \, \mbox{the rotation group}
\end{equation*}
and its consequence
\begin{equation}
U_F(R)| pJ(lS)M_J \rangle = | pJ(lS)M_J^{\prime} \rangle D_{M_J^{\prime}M_J}^{(J)}(R)   
\end{equation}
for infinitisemal rotations.
With the help of \eqref{(B.17)} it is easily seen that
\begin{equation}
{\vec{S}}^{\,2}_{ferm} \left| {pJ(lS)M_J } \right\rangle = S(S+1)
\left| {pJ(lS)M_J } \right\rangle    
\end{equation}
Thus we have built up in the CPR the common eigenvectors of
operators $\vec{J}^2$ and $\vec{S}_{ferm}^2$. Probably, one should
note that $S$ is not any eigenvalue of the so--called invariant
(or internal) spin operator $I=(0,\vec I)$ introduced by M. Shirokov \cite{ChShir58}.
While the latter involves internal orbital motion and polarization
contributions, the quantum number $S$, whose values are regulated
by Clebsch--Gordan coefficient in definition \eqref{(B.11)}, characterizes
rather the total spin of the two--nucleon system. Coming to the end, we allow ourselves to write
the parity operator of the fermion (nucleon) field in the CPR (cf. Eq.(15.93) from \cite{BD}):
\begin{equation*}
\mathscr{P}_{ferm} = exp[i \pi_{ferm}],
\end{equation*}
\begin{multline}
\pi_{ferm} = - \frac{\pi}{2} \int d^3p \left[ b_c^{\dag}(\vec p \mu)b_c(\vec p \mu) - b_c^{\dag}(\vec p \mu)b_c( -\vec p \mu) \right. \\
\left. + d_c^{\dag}(\vec p \mu)d_c(\vec p \mu) + d_c^{\dag}(\vec p \mu)d_c( -\vec p \mu) \right]      \label{(B.21)}
\end{multline}
with the relations
\begin{equation*}
\mathscr{P}_{ferm} b_c(\vec p \mu) \mathscr{P}_{ferm}^{-1} = b_c( -\vec p \mu),
\end{equation*}
\begin{equation*}
\mathscr{P}_{ferm} d^{\dag}_c(\vec p \mu) \mathscr{P}_{ferm}^{-1} =  - d^{\dag}_c( -\vec p \mu)
\end{equation*}
that extend the rules \eqref{(B.4)} for the clothed particle operators to such an improper
transformation as the space inversion.

\section{  The Regularized Quasipotentials and their Angular--Momentum Decomposition}

Trying to overcome ultraviolet divergences inherent in solving
equation \eqref{Eq.50}, we will regularize their driving terms by
introducing some cutoff factors. It can be achieved if instead of
Eq.\eqref{Eq.19} one assumes
\begin{multline}
V_b^{reg}(p_1^{\prime} \mu_1^{\prime},p_2^{\prime} \mu_2^{\prime};p_1 \mu_1,p_2 \mu_2)=
\frac{\delta \left( \vec p_1^{\,\prime} + \vec p_2^{\,\prime}  - \vec p_1  - \vec p_2 \right)}
{\sqrt{p_{01}^{\prime}p_{02}^{\prime}p_{01}p_{02}}} \\
v_b^{reg}(p_1^{\prime} \mu_1^{\prime},p_2^{\prime} \mu_2^{\prime};p_1 \mu_1,p_2 \mu_2)     
\end{multline}
omitting for the moment isospin indices, so
\begin{multline}
K_b(NN \rightarrow NN) \rightarrow K_b^{reg}(NN \rightarrow NN) \\
= \int \delta \left( \vec p_1^{\, \prime} + \vec p_2^{\, \prime}  - \vec p_1  - \vec p_2 \right)
\frac{d \vec p_1^{\, \prime}}{p_{01}^{\prime}} \frac{d \vec p_2^{\, \prime}}{p_{02}^{\prime}}
\frac{d \vec p_1}{p_{01}} \frac{d \vec p_2}{p_{02}} \\
\sum\limits_\mu v_b^{reg}(p_1^{\prime} \mu_1^{\prime},p_2^{\prime} \mu_2^{\prime};p_1 \mu_1,p_2 \mu_2) \\
b_c^{\dag}(p_1^{\prime} \mu_1^{\prime}) b_c^{\dag}(p_2^{\prime} \mu_2^{\prime})
b_c(p_1 \mu_1) b_c(p_2 \mu_2)        
\end{multline}
Here the new (regularized) coefficients $v_b^{reg}$ are given by
\begin{equation}
v_b^{reg} = F_b(p_1^{\prime},p_2^{\prime};p_1,p_2) v_b,  \label{(C.3)}
\end{equation}
where the old ones $v_b$ are determined by
Eq.\eqref{Eq.20}--\eqref{Eq.22} and empirical cutoff functions
$F_b$ \footnote{We do not consider the functions depending on
nucleon polarizations} should not violate the known symmetries of
interactions. In particular, if one writes
\begin{equation}
K_b^{reg}(NN \rightarrow NN) = \int K_b^{reg}(\vec x) d \vec x    
\end{equation}
with
\begin{multline}
K_b^{reg}(\vec x) = \frac{1}{(2\pi)^3} \int
\frac{d \vec p_1^{\, \prime}}{p_{01}^{\prime}} \frac{d \vec p_2^{\, \prime}}{p_{02}^{\prime}}
\frac{d \vec p_1}{p_{01}} \frac{d \vec p_2}{p_{02}} d \vec x
e^{i\left( \vec p_1^{\, \prime} + \vec p_2^{\, \prime}  - \vec p_1  - \vec p_2 \right) \vec x} \\
\sum\limits_\mu v_b^{reg}(p_1^{\prime} \mu_1^{\prime},p_2^{\prime} \mu_2^{\prime};p_1 \mu_1,p_2 \mu_2) \\
b_c(p_1^{\prime} \mu_1^{\prime}) b_c(p_2^{\prime} \mu_2^{\prime})
b_c^{\dag}(p_1 \mu_1) b_c^{\dag}(p_2 \mu_2)    
\end{multline}
then the RI implies the property of the operator
\begin{equation}
K_b^{reg}(x) \equiv \exp \left( iK_Ft \right) K_b^{reg}(\vec x) \exp \left( -iK_Ft \right)   
\end{equation}
to be a scalar, viz.,
\begin{equation}
U_F(\Lambda,\lambda)K_b^{reg}(x)U_F^{-1}(\Lambda,\lambda) = K_b^{reg}(\Lambda x + \lambda)    
\end{equation}
But accordingly \eqref{(B.4)} it imposes the following restrictions
\begin{multline}
D_{\eta_1^{\prime}\mu_1^{\prime}}^{\left( \frac{1}{2} \right)} \left( W(\Lambda, p_1^{\prime})\right)
D_{\eta_2^{\prime}\mu_2^{\prime}}^{\left( \frac{1}{2} \right)} \left( W(\Lambda, p_2^{\prime})\right) \\
\shoveright{ \times D_{\eta_1\mu_1}^{\left( \frac{1}{2} \right)\ast} \left( W(\Lambda, p_1)\right)
D_{\eta_2\mu_2}^{\left( \frac{1}{2} \right)\ast} \left( W(\Lambda, p_2)\right) } \\
\times v_b^{reg}(p_1^{\prime} \mu_1^{\prime},p_2^{\prime} \mu_2^{\prime};p_1 \mu_1,p_2 \mu_2) = \\
v_b^{reg}(\Lambda p_1^{\prime} \eta_1^{\prime},\Lambda p_2^{\prime} \eta_2^{\prime};\Lambda p_1 \eta_1,\Lambda p_2 \eta_2)
\label{(C.8)}
\end{multline}
to the coefficients $v_b^{reg}$. In this connection, before going on, one needs to verify that the old (non--regularized) ones
satisfy relation \eqref{(C.8)} themselves. It can be done with the help of the property
\begin{equation*}
S(\Lambda) u(p\mu) = D_{\mu' \mu}^{(1/2)}(W(\Lambda,p)) u(\Lambda p\mu'),
\end{equation*}
where $S(\Lambda) = exp\left[-\frac{1}{2}\omega_{\mu \nu} \sigma^{\mu \nu} \right]$ the matrix of the nonunitary representation
$\Lambda \rightarrow S(\Lambda)$ in the space of spinor indices. Indeed, recalling more the relations
\begin{equation*}
S(\Lambda)^{-1} \gamma^{\rho} S(\Lambda) = \gamma^{\mu} \Lambda^{\rho}_{\,\, \mu},
\end{equation*}
one can easily seen that the quantities $v_b(p_1^{\prime} \mu_1^{\prime},p_2^{\prime} \mu_2^{\prime};p_1 \mu_1,p_2 \mu_2) = v_b(1',2';1,2)$
by Eqs.(20)--(22) obey Eq.\eqref{(C.8)}.
Also, let us remind (see, e.g., \cite{MelShe}) that for a
Lorentz boost $\Lambda = L(\vec v)$ with the velocity $\vec v$, the Wigner
transformation $W(\Lambda,p)$ is the rotation about the $\vec v \times \vec p$ --direction
by an angle $\psi$, which can be represented as
\begin{equation*}
m \left( 1+\gamma+\frac{p_0^{\star}}{m}+\frac{p_0}{m} \right) \tan
\frac{\psi}{2} = \gamma \left| \vec v \times \vec p \right|,
\end{equation*}
where $p_0^{\star}$ the zeroth component of the nucleon momentum
$p^{\star}=(p_0^{\star},\vec p^{\, \star}) = L(\vec v)p$ in the
moving frame and $\gamma$ the corresponding Lorentz factor. As
noted, $W(R,p)=R$ for a pure rotation $R$.

Now, keeping in mind the relation \eqref{(C.3)} we need to deal with a Lorentz--invariant cutoff,
\begin{equation}
F_b(p_1^{\prime},p_2^{\prime};p_1,p_2) = F_b(\Lambda p_1^{\prime},\Lambda p_2^{\prime};\Lambda p_1,\Lambda p_2)   
\end{equation}
in our model regularization,
\begin{multline}
V_b^{reg}(p_1^{\prime} \mu_1^{\prime},p_2^{\prime} \mu_2^{\prime};p_1 \mu_1,p_2 \mu_2)=
\frac{\delta \left( \vec p_1^{\,\prime} + \vec p_2^{\,\prime}  - \vec p_1  - \vec p_2 \right)}
{\sqrt{p_{01}^{\prime}p_{02}^{\prime}p_{01}p_{02}}} \\
F_b(p_1^{\prime},p_2^{\prime};p_1,p_2)
v_b(p_1^{\prime} \mu_1^{\prime},p_2^{\prime} \mu_2^{\prime};p_1 \mu_1,p_2 \mu_2)    
\end{multline}
Further, assuming that $F_b$ depends on the two invariants $(p_1 - p_1^{\prime})^2$ and $(p_2 - p_2^{\prime})^2$,
\begin{equation*}
F_b(p_1^{\prime},p_2^{\prime};p_1,p_2) =
f_b\left( (p_1 - p_1^{\prime})^2,(p_2 - p_2^{\prime})^2 \right)
\end{equation*}
we have on the momentum shell,
\begin{equation*}
(p_1 - p_1^{\prime})^2 = (E_1 - E_1^{\prime})^2 - (\vec p_1 - \vec p_1^{\, \prime})^2
\end{equation*}
\begin{equation*}
(p_2 - p_2^{\prime})^2 = (E_2 - E_2^{\prime})^2 - (\vec p_1 - \vec p_1^{\, \prime})^2
\end{equation*}
so in the c.m.s. one has to deal with the function
$f_b[(p-p^{\prime})^2]$ ($F^{2}_b\left[ (p^{\prime}-p)^2 \right]$
in the main text) of one invariant $(p-p^{\prime})^2=(E_{\vec
p}-E_{\vec p^{\, \prime}})^2-(\vec p - \vec p^{\, \prime})^2$.

Of course, a similar regularization could be implemented if the
vertices
\begin{equation*}
\bar u(p' \mu') \Gamma_b(k) u(p \mu)
\end{equation*}
(see, e.g., Eq.(92)) would be at the beginning modified
by
introducing cutoff factors $g_{++}^b(p',p)$ with the bosons and
fermions on their mass shells $k^2 = m_b^2$, $p'^2 = p^2 = m^2$
and the momentum conservation $\vec k = \vec p' - \vec p$.
Neglecting a possible dependence of such factors on particle
polarizations, we preserve the subscript $++$ keeping in mind a
more symmetrical consideration, where some factors
$g_{\varepsilon' \varepsilon}^b(p',p)$ could be introduced with the
proper energy--sign labels $\varepsilon'$ and $\varepsilon$ for
separate three--legs contributions to a given boson--fermion
Yukawa--type interaction. Then the corresponding regularized
quasipotentials would be contained factors
\begin{equation}
F_b(p_1^{\prime},p_2^{\prime};p_1,p_2) =
g_{++}^b(p_1',p_1)g_{++}^b(p_2,p_2')
\end{equation}
in the quadrilinear terms.

As noted below Eq.(66) our choice of the phenomenological cutoff $F_b[(p'-p)^2]$ has been prompted by those investigations \cite{MachHolElst87} of the Bonn group. Of course, recent developments in studying the structure of meson--baryon interaction vertices are of great interest. Among certain achievements in this area we find a microscopic derivation of the strong $\pi NN$ and $\pi N \Delta$ FFs within the quark model developed by the Graz group (see, e.g., \cite{Graz}). Being free of any phenomenological input (fit parameters) the corresponding prediction for the FF $G_{\pi NN}$ (e.g., in an analytic form put forward in \cite{Graz}) could be employed as the bare factor $g^{\pi}_{++}(p',p)$. The latter has the property $g^{\pi}_{++}(\Lambda p',\Lambda p)=g^{\pi}_{++}(p',p)$, i.e., it should be dependent upon the Lorentz scalar $p'p$ or $(p'-p)^2$ to fulfill condition (86). Other vertex cutoffs $g^{\pi}_{+-}$, $g^{\pi}_{-+}$ and $g^{\pi}_{--}$ are also desirable to be introduced on the same physical footing.

By the way, along with properties $g^b_{\pm \pm}(\Lambda p',
\Lambda p) =g^b_{\pm \pm}(p',p)$ condition (86) means that the
crossed cutoffs $g^b_{+-}(p',p)$ and $g^b_{-+}(p',p)$ should be
some functions of the four--product $p'_{-}p=p'p_{-}$. In turn,
other symmetries, mentioned below Eq.(69), yield the following
links $g^b_{\varepsilon ' \varepsilon}(p',p)=g^b_{\varepsilon '
\varepsilon}(p,p')$, $g^b_{++}(p',p)=g^b_{--}(p',p)$,
$g^b_{+-}(p',p)=g^b_{-+}(p,p')$ for the real functions. All the
things are extremely important for constructing a relativistic
nonlocal QFT, where the boosts operators are determined as
elements of the Lie algebra of the Poincar\'{e} group. It will be
presented somewhere else.

In addition, let us address the matrix elements $\langle p' \mid j_b(0) \mid p \rangle$  that are closely connected with the FFs in question (see, e.g., \cite{Gas66}) since we handle the corresponding baryon current density $j_b(x)$ at $x=0$ sandwiched between physical (clothed) one--nucleon states. Such matrix elements might be evaluated in terms of the cutoffs and other physical inputs using some idea from \cite{SheShi00} (cf. the clothed particle representation of a current operator therein). From the constructive point of view, it means the use of definitions $\mid p \rangle=b_c^{\dag}(p \mu)\mid \Omega \rangle$ and $\mid p' \rangle=b_c^{\dag}(p' \mu')\mid \Omega \rangle$, adopted at the beginning of Appendix B, and the similarity transformation
\begin{multline*}
j_b(0)=e^{R(\alpha_c)}j_b^c(0) e^{-R(\alpha_c)} \\
= j_b^c(0) + \left[ R(\alpha_c) ,j_b^c(0) \right] + \frac{1}{2!}
\left[R(\alpha_c) ,\left[R(\alpha_c) ,j_b^c(0) \right] \right] \\
+\frac{1}{3!} \left[ R(\alpha_c) , \left[ R(\alpha_c) ,\left[
R(\alpha_c) ,j_b^c(0) \right] \right] \right] + \cdots,
\end{multline*}
where $j_b^c(0)$ is the same current density but expressed through
the clothed operators. The nonperturbative expansion in the
commutators gives an opportunity for a systematic evaluation of
corrections to matrix elements
\begin{equation*} \langle \Omega
\mid b_c j_b^c(0) b_c^{\dag} \mid \Omega \rangle = \langle
\Omega_0 \mid b j_b(0) b^{\dag} \mid \Omega_0 \rangle.
\end{equation*}
Some simplifications originate from the well--known
fact that similar expectations of the commutators that involve odd
number of meson operators are equal to zero.

In general, one can elaborate a recursive procedure of calculations, like that by Kharkov-Padova group, for manipulations with the multiple commutators $[V]^n$ ($n=2,3,\cdots$) (see \cite{KorCanShe07}). Doing so, one can find corrections to formula (125) obtained from the commutator $[V]^1=[R,V]$. This work is in progress.

After this prelude we note that the partial--wave matrix elements of interest are defined by
\begin{multline}
\bar V_{l^{\prime}l}^{JS}=\sum\limits_b {^{b}} \bar V_{l^{\prime}l}^{JS} = \frac{1}{2J+1} \sum\limits_b
\int d \hat {\vec p^{\prime}} \int d \hat {\vec p} \,\, Y_{l^{\prime}m_l^{\prime}}^{\ast} (\hat {\vec p^{\prime}})
Y_{lm_l} (\hat {\vec p}) \\
\left( l^{\prime}m_l^{\prime}SM_S^{\prime} \left|JM_J\right. \right)
\left( lm_lSM_S \left|JM_J\right. \right)
\langle \left. \vec p^{\, \prime} SM_S^{\prime} \right| \bar V_b \left| \vec p SM_S \right. \rangle    
\end{multline}
with
\begin{multline}
\langle \left. \vec p^{\, \prime} SM_S^{\prime} \right| \bar V_b \left| \vec p SM_S \right. \rangle = \\
\bar V_{dir}^{b}(\vec p^{\, \prime} SM_S^{\prime},\vec p SM_S) -
\bar V_{exc}^{b}(\vec p^{\, \prime} SM_S^{\prime},\vec p SM_S),    
\end{multline}
\begin{multline}
\bar V_{dir(exc)}^{b}(\vec p^{\, \prime} SM_S^{\prime},\vec p SM_S) =
- \frac{1}{2(2\pi)^3} \frac{m^2}{E_{p^{\prime}}E_p} \\
\left( \frac{1}{2} \mu_1^{\prime} \frac{1}{2} \mu_2^{\prime} \left|SM_S^{\prime}\right. \right)
\left( \frac{1}{2} \mu_1 \frac{1}{2} \mu_2 \left|SM_S\right. \right)
v_b^{dir(exc)}(\vec p^{\, \prime} \mu_1^{\prime} \mu_2^{\prime},\vec p \mu_1 \mu_2),   \label{(C.13)}
\end{multline}
where we have employed formula (45) and property (48). \\
In turn, the matrices $v_b^{dir}(\vec p^{\, \prime} \mu_1^{\prime} \mu_2^{\prime},\vec p \mu_1 \mu_2)$ can be represented as
\begin{equation}
v_b^{dir} = \Gamma_b(\vec p^{\, \prime} \mu_1^{\prime};\vec p
\mu_1)D_b(p^{\prime},p)\Gamma_b(-\vec p^{\, \prime}
\mu_2^{\prime};-\vec p \mu_2),
\end{equation}
\begin{equation}
D_b(p^{\prime},p)=\frac{F^{2}_b\left[ (p^{\prime}-p)^2 \right]}{(p^{\prime}-p)^2-m_b^2}.   
\end{equation}
Recall that
\begin{equation*}
v_b^{exc}(\vec p^{\, \prime} \mu_1^{\prime} \mu_2^{\prime},\vec p \mu_1 \mu_2) =
v_b^{dir}(\vec p^{\, \prime} \mu_1^{\prime} \mu_2^{\prime},-\vec p \mu_2 \mu_1),
\end{equation*}
so
\begin{equation}
\bar V_{exc}^{b}(\vec p^{\, \prime} SM_S^{\prime},\vec p SM_S) =
(-1)^{S+1}\bar V_{dir}^{b}(\vec p^{\, \prime} SM_S^{\prime},- \vec p SM_S).    
\end{equation}
Further, taking into account the completeness of the matrices ${1,\vec \sigma}$ in $2\times2$ space,
the non--regularized vertices from Eqs.\eqref{Eq.20}--\eqref{Eq.22} can be written as
\begin{equation*}
\Gamma_b(\vec p^{\, \prime} \mu^{\prime};\vec p \mu) = A_b(\vec p^{\, \prime} ;\vec p )\delta_{\mu^{\prime} \mu}
+ \mbox{ a linear functional of } \vec \sigma
\end{equation*}
Now, to get the matrix $\langle \left. \vec p^{\, \prime} SM_S^{\prime} \right| \bar V_b \left| \vec p SM_S \right. \rangle$ we could do all summations
in formula \eqref{(C.13)} over $\mu$ projections directly. However, we prefer the following way putting formally
\begin{equation*}
\langle \mu_1^{\prime} | \vec \sigma | \mu_1 \rangle = \langle \mu_1^{\prime} | \hat {\vec \sigma}(1) | \mu_1 \rangle \,\,\,\, \mbox{and} \,\,\,\,
\langle \mu_2^{\prime} | \vec \sigma | \mu_2 \rangle = \langle \mu_2^{\prime} | \hat {\vec \sigma}(2) | \mu_2 \rangle
\end{equation*}
to obtain
\begin{equation}
\bar V_{dir}^b (\vec p^{\, \prime} SM_S^{\prime}, \vec p SM_S)  =
\langle \left. SM_S^{\prime} \right| G_b(\vec p^{\, \prime},\vec p;\hat {\vec \sigma}(1),\hat {\vec \sigma}(2))\left| SM_S \right. \rangle
\label{(C.17)}
\end{equation}
with the $SM_S$  eigenvalue equations,
\begin{equation*}
\hat {\vec S}^2 | S M_S \rangle = S(S+1)| S M_S \rangle
\end{equation*}
\begin{equation*}
\hat S_3 | S M_S \rangle = M_S| S M_S \rangle,
\end{equation*}
where $\hat {\vec S} = \frac{1}{2}\left[\hat {\vec \sigma}(1) +\hat {\vec \sigma}(2) \right]$. The operators
$G_b$ can be expressed through the operator $\hat {\vec S}$ with the help of the relations
\begin{equation*}
\hat {\vec \sigma}(1) \cdot \vec n \,\,\, \hat{\vec \sigma}(2) \cdot \vec n = 2\left( \hat{\vec S} \cdot \vec n \right)^2 - \vec n^2
\end{equation*}
\begin{equation*}
\hat{\vec \sigma}(1) \times \vec n \,\,\, \hat{\vec \sigma}(2) \times \vec n = \hat{\vec \sigma}(1) \cdot \hat{\vec \sigma}(2) \,\,\, \vec n^2 -
\hat{\vec \sigma}(1) \cdot \vec n \,\,\, \hat{\vec \sigma}(2) \cdot \vec n 
\end{equation*}
for any vector $\vec n$. \\
As a result, we find with the models \eqref{Eq.20}--\eqref{Eq.22}
\begin{equation}
G_s = \frac{g_s^2}{2(2\pi)^3}CD_s(p',p)\left\{ U_1^2 + \vec V_3^2 - 2(\vec V_3 \vec S) \left[ iU_1 + (\vec V_3 \vec S) \right] \right\},
\end{equation}
\begin{equation}
G_{ps} = \frac{g_{ps}^2}{2(2\pi)^3}CD_{ps}(p',p)\left\{ \vec V_1^2 - 2(\vec V_1 \vec S)^2 \right\},
\end{equation}
\begin{equation}
G_{\rm{v}} = G_{\rm{v}}^{vv} + G_{\rm{v}}^{vt} + G_{\rm{v}}^{tt}, 
\end{equation}
\begin{equation}
G_{\rm{v}}^{vv} = \frac{g_{\rm{v}}^2}{2(2\pi)^3}CD_{\rm{v}}(p^{\prime},p)G_1, 
\end{equation}
\begin{multline}
G_{\rm{v}}^{vt} = \frac{1}{2(2\pi)^3}\frac{f_{\rm{v}}g_{\rm{v}}}{2m}CD_{\rm{v}}(p',p)\left\{ 4mG_1 \right.  \\
\left. - 2\left[ (E_{p'}+E_p)G_2 + G_3 + (E_{p'}-E_p)G_4 \right] \right\}, 
\end{multline}
\begin{multline}
G_{\rm{v}}^{tt} = \frac{1}{2(2\pi)^3}\frac{f_{\rm{v}}^2}{4m^2}CD_{\rm{v}}(p',p)\left\{ 4m^2G_1 \right. \\
- 4m\left[ (E_{p'}+E_p)G_2 + G_3 + (E_{p'}-E_p)G_4 \right] \\
\left. + G_5 + 2(E_{p'}-E_p)G_6  \right\},  
\end{multline}
with
\begin{multline*}
G_1 = U_2^2 + \vec V_3^2 + \vec V_2^2 -2(\vec S^2-1)\vec V_1^2 \\
+2(\vec V_3 \vec S) \left[ iU_2 - (\vec V_3 \vec S) \right] +2(\vec V_1 \vec S)^2 + 2i(\vec V_2 \times \vec V_1)\vec S,
\end{multline*}
\begin{equation*}
G_2 = U_1U_2 - \vec V_3^2 + 2(\vec V_3 \vec S) \left[ i(U_1-U_2) + 2(\vec V_3 \vec S) \right],
\end{equation*}
\begin{multline*}
G_3 = U_1\vec B \vec V_2 + i(\vec B \vec V_2)(\vec V_3 \vec S) + iU_1(\vec B \times \vec V_1) \vec S \\
- (\vec B \times \vec V_1) \left[ 2\vec S (\vec V_3 \vec S) - \vec V_3 - i (\vec V_3 \times \vec S)  \right],
\end{multline*}
\begin{equation*}
G_4 = (5-2\vec S^2)\vec V_1 \vec V_2 + (\vec V_1 \vec S)(\vec V_2 \vec S) + (\vec V_2 \vec S)(\vec V_1 \vec S),
\end{equation*}
\begin{equation*}
G_5 = \vec B^2 \left\{ U_1^2 + \vec V_3^2 + 2(\vec V_3 \vec S) \left[ iU_1 - (\vec V_3 \vec S) \right] \right\},
\end{equation*}
\begin{multline*}
G_6 = U_1\vec B \vec V_1 + i(\vec B \vec V_1)(\vec V_3 \vec S) + iU_1(\vec B \times \vec V_2) \vec S \\
- (\vec B \times \vec V_2) \left[ 2\vec S (\vec V_3 \vec S) - \vec V_3 - i (\vec V_3 \times \vec S)  \right],
\end{multline*}
where we use the notations
\begin{equation*}
U_1(\vec p^{\, \prime},\vec p)=(E_{p'}+m)(E_p+m) - \vec p^{\, \prime}\vec p \equiv U_1,
\end{equation*}
\begin{equation*}
U_2(\vec p^{\, \prime},\vec p)=(E_{p'}+m)(E_p+m) + \vec p^{\, \prime}\vec p \equiv U_2,
\end{equation*}
\begin{equation*}
\vec V_1(\vec p^{\, \prime},\vec p)=\vec p (E_{p'}+m) - \vec p^{\, \prime} (E_{p'}+m) \equiv \vec V_1,
\end{equation*}
\begin{equation*}
\vec V_2(\vec p^{\, \prime},\vec p)=\vec p (E_{p'}+m) + \vec p^{\, \prime} (E_p+m) \equiv \vec V_2,
\end{equation*}
\begin{equation*}
\vec V_3(\vec p^{\, \prime},\vec p)= \vec p^{\, \prime} \times \vec p \equiv \vec V_3, {\,\,\,\,\,\,\,\,}
\vec B(\vec p^{\, \prime},\vec p)= \vec p^{\, \prime} + \vec p \equiv \vec B
\end{equation*}
and
\begin{equation*}
C = \left[ 4E_{p'}E_{p'}(E_{p'}+m)(E_p+m) \right]^{-1}.
\end{equation*}
These expressions were used by us to evaluate
the matrix elements of interest,
\begin{multline}
{^{b}} \bar V_{l^{\prime}l}^{JS} = \frac{1}{2J+1}
\int d \hat {\vec p^{\,\prime}} \int d \hat {\vec p} \\
\times \langle \textit{Y}_{JM_J}^{\,l'S} (\hat {\vec p^{\,\prime}})|
G_b(\vec p^{\,\prime},\vec p;\vec S)+(-1)^S G_b(\vec p^{\,\prime},-\vec p;\vec S)  | \textit{Y}_{JM_J}^{\,lS} (\hat {\vec p}) \rangle \\
= \frac{ \left[ 1+(-1)^{S+l}\right] }{2J+1} \int d \hat {\vec p^{\,\prime}} \int d \hat {\vec p} \\
\times \langle \textit{Y}_{JM_J}^{\,l'S} (\hat {\vec p^{\,\prime}})| G_b(\vec p^{\,\prime},\vec p;\vec S)
| \textit{Y}_{JM_J}^{\,lS} (\hat {\vec p}) \rangle ,   \label{(C.24)}
\end{multline}
where
\begin{equation*}
| \textit{Y}_{JM_J}^{\,lS} (\hat {\vec n}) \rangle = Y_{lm_l} (\hat {\vec n})
\left( l^{\prime}m_l^{\prime}SM_S^{\prime} \left|JM_J\right. \right) |SM_S\rangle
\end{equation*}
the so--called spin--angular states.

A simple extension to the states with the isospin $T$ yields the
factor $I_b(T)\left[ 1-(-1)^{S+l+T} \right]$ instead of $\left[
1+(-1)^{S+l} \right]$ in the r.h.s. of Eq.\eqref{(C.24)}. Its
appearance results in the well--known selection rule for the
nucleon--nucleon scattering. Here
\begin{equation}
I_b(T) = \left\{ {\begin{array}{*{20}c}
   1 & \mbox{for neutral bosons,}  \\
   {2T(T + 1) - 3} & \mbox{for charged bosons}  \\
\end{array}} \right.
.  
\end{equation}
The operators $G_b(\vec p^{\,\prime},\vec p;\vec S)$ depend on the scalars $({\vec p^{\,\prime}}{\vec p})$,
$(\vec S {\vec p^{\,\prime}})^2$, $(\vec S {\vec p})^2$, $(\vec S {\vec p^{\,\prime}})(\vec S {\vec p})$,
$(\vec S {\vec p})(\vec S {\vec p^{\,\prime}})$, so all we need is to calculate the following integrals:
\begin{multline*}
{^b}I_{l'l}^{JS}(p',p) \\
=\int d \hat {\vec p^{\,\prime}} \int d \hat {\vec p}
\langle \textit{Y}_{JM_J}^{\,l'S} (\hat{\vec p^{\,\prime}})| D_b(p',p)I(\vec p^{\,\prime},\vec p;\vec S)
| \textit{Y}_{JM_J}^{\,lS} (\hat {\vec p}) \rangle,
\end{multline*}
where $I(\vec p^{\,\prime},\vec p;\vec S)$ is a polynomial of these scalars.

After a lengthy calculation we arrive to our working formulae. In
particular, in case of the tensor--tensor interaction in the
$\rho$--exchange channel we have for uncoupled waves
\begin{multline}
^{\rho,tt}V^{J0}_{JJ}(p',p) = - \frac{f^2_{\rm{v}}}{4\pi} \frac{1}{8\pi^2E'Em^2} \\
\left\{ \left[ p^2p^{\prime \, 2} + (p^2+p^{\prime \, 2})(E'E-2m^2) + 6m^2(E'E-m^2) \right]{^{\rho}}\widetilde{Q}_J(p',p) \right. \\
- pp'(p^2+p^{\prime \, 2}+4m^2){^{\rho}}\widetilde{Q}^{(1)}_J(p',p) - p^2p^{\prime \, 2}{^{\rho}}\widetilde{Q}^{(4)}_J(p',p) \\
\left. - (E'-E)^2 \left[ (E'E-5m^2){^{\rho}}\widetilde{Q}_J(p',p) - pp'{^{\rho}}\widetilde{Q}^{(1)}_J(p',p) \right] \right\},
\end{multline}
\begin{multline}
^{\rho,tt}V^{J1}_{JJ}(p',p) = - \frac{f^2_{\rm{v}}}{4\pi} \frac{1}{8\pi^2E'Em^2} \\
\left\{ \left[ p^2p^{\prime \, 2} + (p^2+p^{\prime \, 2})(E'E-2m^2) + 2m^2(E'E-m^2) \right]{^{\rho}}\widetilde{Q}_J(p',p) \right. \\
+ pp'(E'E+m^2){^{\rho}}\widetilde{Q}^{(1)}_J(p',p) \\
- pp'(E'E+m^2p^2+p^{\prime \, 2}){^{\rho}}\widetilde{Q}^{(2)}_J(p',p) - p^2p^{\prime \, 2}{^{\rho}}\widetilde{Q}^{(5)}_J(p',p) \\
\left. - (E'-E)^2 \left[ (E'E-m^2){^{\rho}}\widetilde{Q}_J(p',p) - pp'{^{\rho}}\widetilde{Q}^{(2)}_J(p',p) \right] \right\},
\end{multline}
with
\begin{equation}
{^{\rho}}\widetilde{Q}^{(1)}_J(p',p) = \frac{1}{2J+1}\left\{ J{^{\rho}}\widetilde{Q}_{J-1}(p',p) +
(J+1){^{\rho}}\widetilde{Q}_{J+1}(p',p) \right\},
\end{equation}
\begin{equation}
{^{\rho}}\widetilde{Q}^{(2)}_J(p',p) = \frac{1}{2J+1}\left\{ (J+1){^{\rho}}\widetilde{Q}_{J-1}(p',p) +
J{^{\rho}}\widetilde{Q}_{J+1}(p',p) \right\},
\end{equation}
\begin{multline}
{^{\rho}}\widetilde{Q}^{(4)}_J(p',p) = \frac{1}{2J+1}\left\{ \frac{J(J-1)}{2J-1}{^{\rho}}\widetilde{Q}_{J-2}(p',p)  \right. \\
\left. \frac{2J^2(2J+3)-1}{(2J-1)(2J+3)}{^{\rho}}\widetilde{Q}_J(p',p)
+ \frac{(J+1)(J+2)}{2J+3}{^{\rho}}\widetilde{Q}_{J+2}(p',p)  \right\},
\end{multline}
\begin{multline}
{^{\rho}}\widetilde{Q}^{(5)}_J(p',p) = \frac{1}{2J+1}\left\{ \frac{J^2-1}{2J-1}{^{\rho}}\widetilde{Q}_{J-2}(p',p)  \right. \\
\left. \frac{2J(J+1)(2J+1)}{(2J-1)(2J+3)}{^{\rho}}\widetilde{Q}_J(p',p)
+ \frac{J(J+2)}{2J+3}{^{\rho}}\widetilde{Q}_{J+2}(p',p)  \right\}.
\end{multline}
In these formulae
\begin{equation*}
{^{b}}\widetilde{Q}_n(p',p) = 2\pi \int \limits_{-1}^1
d(\cos\theta)P_n(\cos\theta)D_b(p',p),
\end{equation*}
where $P_n(\cos\theta)$ is the Legendre polynomial.
Using the Neumann integral representation for the Legendre function of second kind
$Q_n(x)$ one can write for any $n_b$
\begin{multline*}
\widetilde{Q}_n^{b}(p',p) = 2\pi \int \limits_{-1}^1 d(\cos\theta) \frac{P_n(\cos\theta)}{(p'-p)^2 - m_b^2}
\left[ \frac{\Lambda_b^2 - m_b^2}{(p'-p)^2 - \Lambda_b^2} \right]^{2n_b} \\
=-4\pi\frac{\left[ \Lambda_b^2 - m_b^2 \right]^{2n_b}}{(2p'p)^{2n_b+1}} \left\{ \frac{Q_n(x)}{(y-x)^{n_b}}  \right. \\
\left. + \sum \limits_{m=0}^{2n_b-1}(-1)^{m+1}\frac{1}{m!(y-x)^{n_b-m}}\frac{d^m}{dy^m}Q_n(y) \right\},
\end{multline*}
\begin{equation*}
x = \frac{p^2 + p^{\prime 2}+m_b^2 - (E_{p'}-E_{p})^2}{2p'p},
\end{equation*}
\begin{equation*}
y = \frac{p^2 + p^{\prime 2}+\Lambda_b^2 - (E_{p'}-E_{p})^2}{2p'p}
\end{equation*}
and
\begin{equation*}
\widetilde{Q}_{n<0}^{b}(p',p) \equiv 0.
\end{equation*}.



\end{document}